\documentclass[reprint,amsmath,amssymb, aps]{revtex4-2}
\usepackage{graphicx}
\usepackage{dcolumn}
\usepackage{bm}
\usepackage{hyperref}

\def\mathu{{\bm{\mathcal{U}}}}
\def\mathp{{\bm{\mathcal{P}}}}
\def\kB{k_{\rm B}}
\def\avg#1{{\langle #1\rangle }}
\def\ba{{\bf a}}
\def\bb{{\bf b}}
\def\bc{{\bf c}}
\def\bu{{\bf u}}
\def\br{{\bf r}}
\def\bR{{\bf R}}
\def\bp{{\bf p}}
\def\dya#1{|#1\rangle\langle#1|}
\def\ket#1{|#1\rangle }

\begin{document}
\title{Vibrational entropy of crystalline solids from covariance of atomic displacements}
\author{Yang Huang}
\affiliation{
  Physics Department, Carnegie Mellon University.
}
\author{Michael Widom}
\affiliation{
  Physics Department, Carnegie Mellon University.
}

\date{\today}

\begin{abstract}
The vibrational entropy of a solid at finite temperature is investigated from the perspective of information theory.  Ab initio molecular dynamics (AIMD) simulations generate ensembles of atomic configurations at finite temperature from which we obtain the $N$-body distribution of atomic displacements, $\rho_N$. We calculate the information-theoretic entropy from the expectation value of $\ln{\rho_N}$. At a first level of approximation, treating individual atomic displacements independently, our method may be applied using Debye-Waller B-factors, allowing diffraction experiments to obtain an upper bound on the thermodynamic entropy. At the next level of approximation we correct the overestimation through inclusion of displacement covariances. We apply this approach to elemental body-centered cubic sodium and face-centered cubic aluminum, showing good agreement with experimental values above the Debye temperatures of the metals. Below the Debye temperatures we extract an effective vibrational density of states from eigenvalues of the covariance matrix, and then evaluate the entropy quantum mechanically, again yielding good agreement with experiment down to low temperatures. Our method readily generalizes to complex solids, as we demonstrate for a high entropy alloy. Further, our method applies in cases where the quasiharmonic approximation fails, as we demonstrate by calculating the HCP/BCC transition in Ti.
\end{abstract}

\maketitle

\section{Introduction}
\label{sec:intro}
The importance of entropy as a component of thermodynamic free energy, together with the difficulty of its calculation, motivates continuing efforts seeking improved computational approaches~\cite{Morris1995,Meirovitch2009,Widom2016,Sutton2020,Nir2020,Nicholson2021}. The entropy is a function of the state of the system, and is in principle determined by the instantaneous values of every degree of freedom. Most computational approaches to entropy calculation do not make explicit use of these values, and instead apply some form of thermodynamic integration to relate the entropy in the state of interest to some reference point of known entropy~\cite{FrenkelSmit,Kastner2005,Grabowski2019}. Our approach recognizes that the entropy equals, in suitable units, the information required to fully specify the state of the system. We capture this information in the form of many-body correlation functions obtained from {\em ab initio} molecular dynamics (AIMD).

Multiple types of excitation contribute to the entropy of a solid. Neglecting correlations among these, we may approximate the entropy as a sum
\begin{equation}
  \label{eq:1}
  S\approx S^v + S^e + S^c + \cdots
\end{equation}
where $S^v$ arises from atomic vibrations~\cite{Fultz2010}, $S^e$ includes electronic excitations, the non-vibrational configurational term $S^c$ incorporates vacancies and chemical species substitutions~\cite{Widom2018}. The additional terms may include magnetism and other effects~\cite{Grabowski2015}. The present paper primarily addresses the vibrational contribution, but for comparison with experiment we must include the electronic entropy. While our initial approach is classical, and intended for applicability at elevated temperatures close to melting, we show how quantum effects can be incorporated to obtain accurate results below the Debye temperature. Additionally, the electronic entropy is intrinsically a quantum mechanical property.

The following section presents our computational methods. The heart of our approach rests on approximating the many-body displacement correlation function as a Gaussian distribution matching the simulated covariance of atomic displacements. We then apply the method to two test cases, face-centered cubic Al and body-centered cubic Na. In each case we compare with standard thermodynamic data. We also show the applicability of a simple approximation based on experimental Debye-Waller factors (thermal B-factors) that allow experimental diffraction measurements to obtain approximate thermodynamic entropies.

\begin{figure*}[htpb]
  \centering
  \includegraphics[height=.35\textwidth]{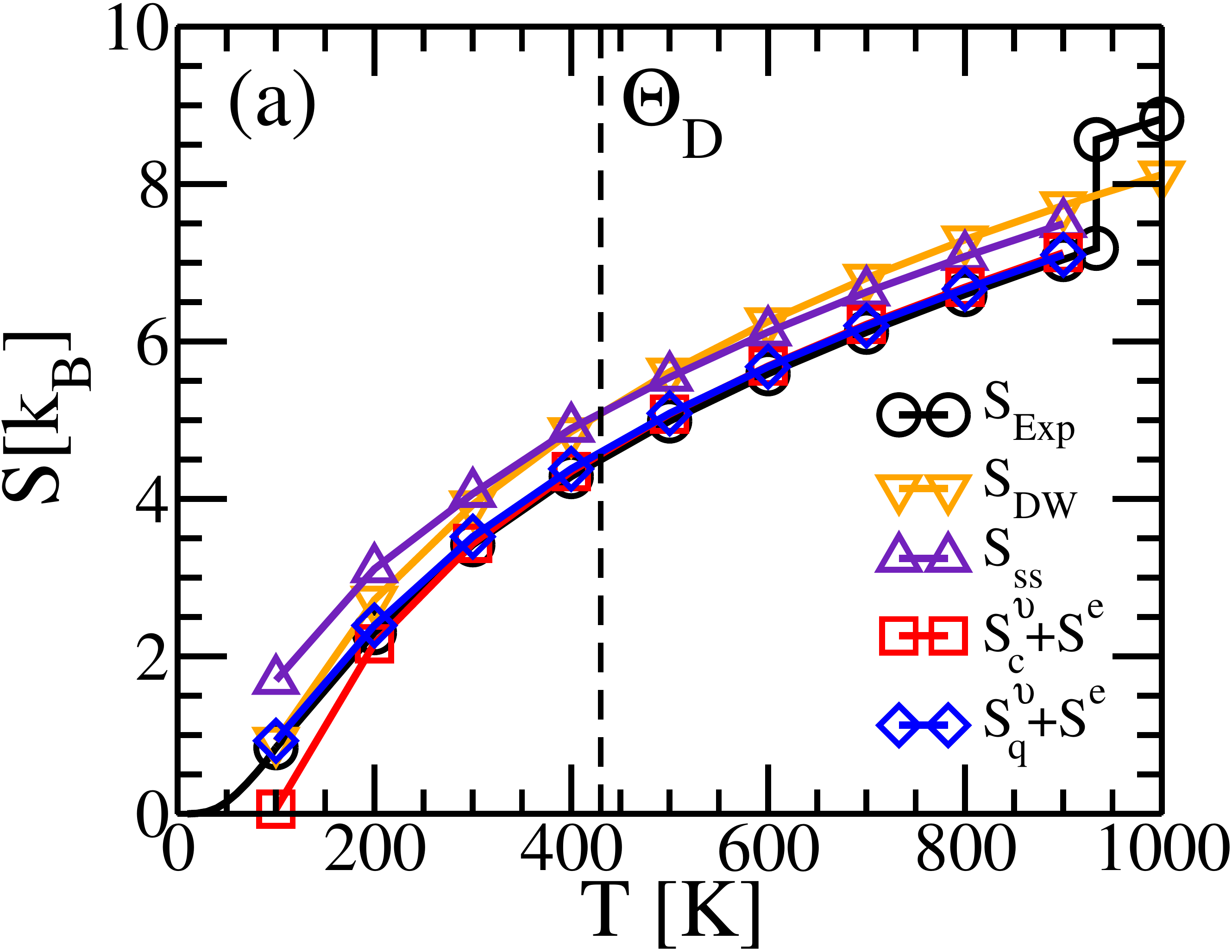}
  \includegraphics[height=.35\textwidth]{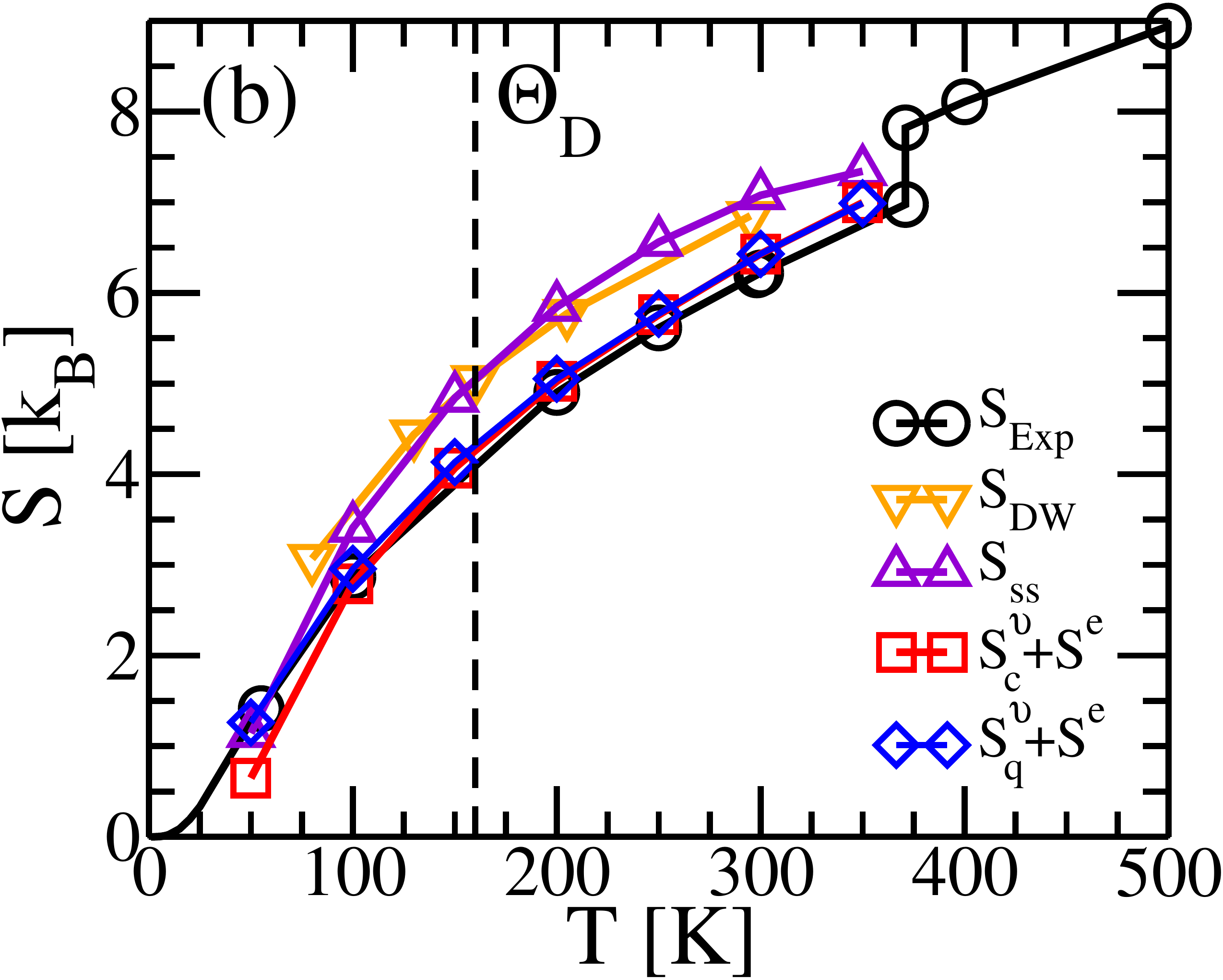}
  \includegraphics[height=.35\textwidth]{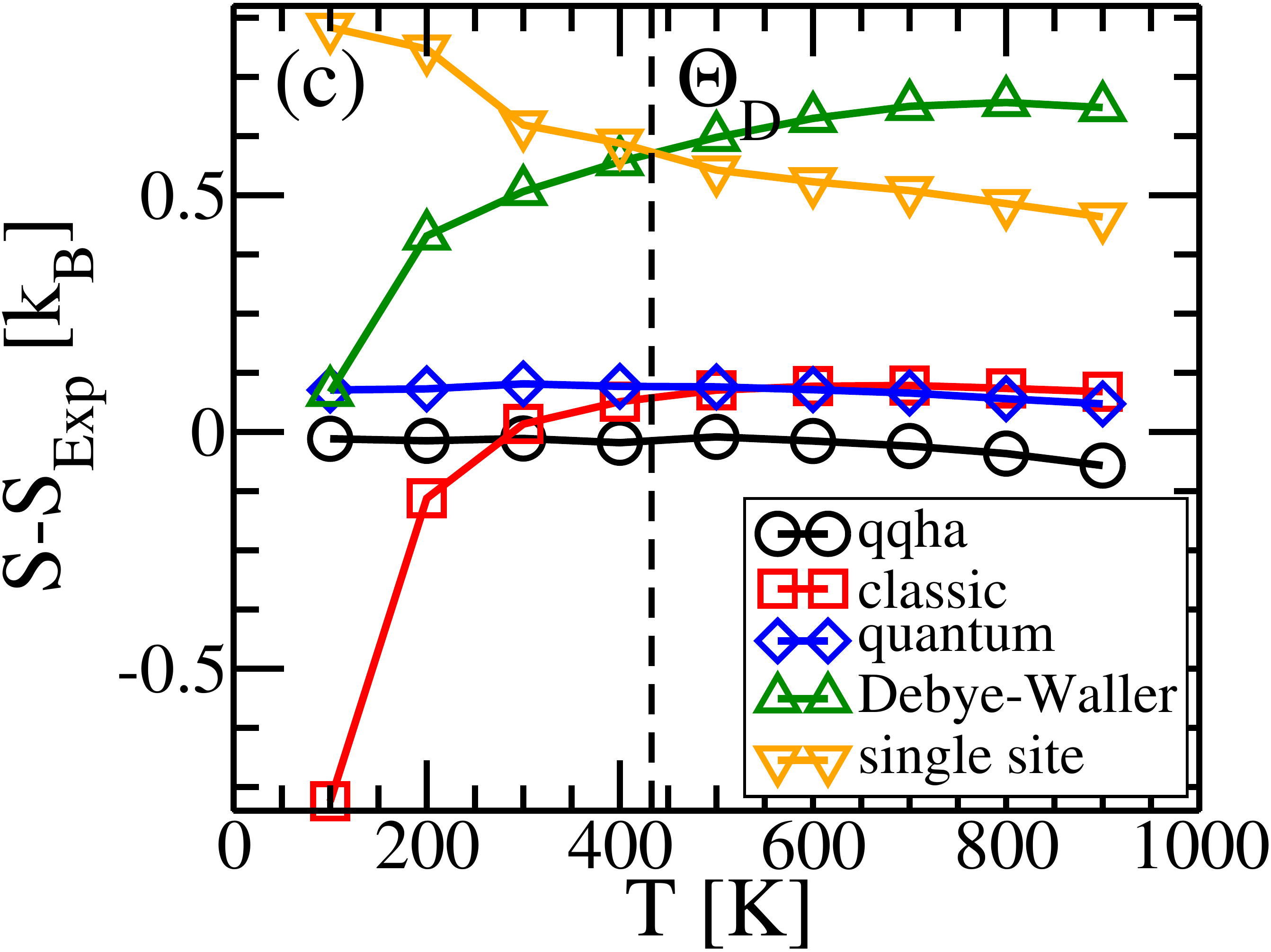}
  \includegraphics[height=.35\textwidth]{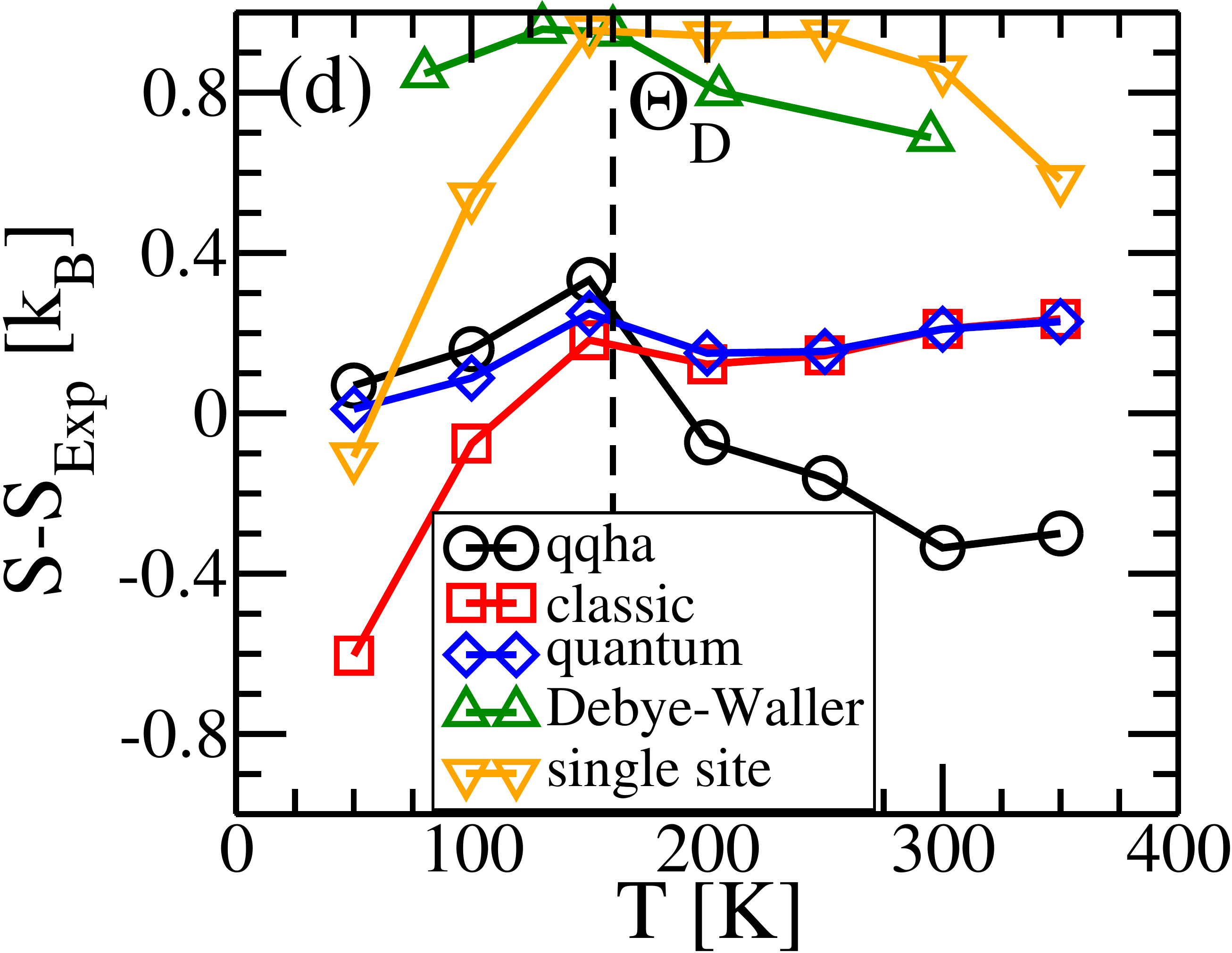}
  \caption{\label{fig:results} Entropies of (a) Al, and (b) Na. Black circles show experimental values from the NIST JANAF Tables~\cite{JANAFAl001,JANAFNA001}. Orange triangles are calculated from Eq.~(\ref{eq:single-site})  using B-factors obtained from~\cite{Nakashima2019,Foust1972}. Red squares add the classical vibrational entropy calculated from Eq.~(\ref{eq:S:cov}) to the electronic entropy calculated from Eq.~(\ref{eq:Se}). Parts (c) and (d) show residuals after subtracting the experimental values. In addition, we show our vibrational quantum model calculated from Eq.~(\ref{eq:quantumS}) using effective vibrational frequencies calculated by Eq.~(\ref{eq:Inverse}), and the quasiharmonic prediction with vibrational frequencies calculated by Phonopy. All calculations are performed at the experimentally determined volumes for each temperature. All curves, except for Debye-Waller and single-site, include electronic entropy.}
\end{figure*}

Our principal results for these test cases are illustrated in Fig.~\ref{fig:results} parts (a) and (b). Notice first that the Debye-Waller factors yield good qualitative results, lying within 1$\kB$ of the experimental values, but remaining consistently high. This is because the Debye-Waller factors treat the individual atomic vibrations independently, and neglect the mutual information contained in displacement correlation functions that must reduce the vibrational entropy~\cite{Morris1995,GaoWidom2018,WidomGao2019,HuangGaoWidom2021,Nicholson2021}. Including the covariances of displacements and electronic entropies (curves labeled classical) improves the agreement, but with negative deviations at low temperatures due to the $\log{T}$ divergence of the classical vibrational entropy.

To overcome the deficiency of classical statistical mechanics at low temperatures, we introduce a quantum version of our method where we interpret eigenvalues of the covariance matrix as effective vibrational frequencies renormalized by anharmonic forces. This reveals a relationship between our method and a different approach based on velocity autocorrelation functions~\cite{Rahman1964,Dickey1969,Lin2003}, although the two approaches differ in important respects. Residuals of several versions of our calculated entropies relative to experiment are plotted in Fig.~\ref{fig:results} (c) and (d), and are further discussed in Section~\ref{sec:TestCases}.

We then apply our method to two examples that are scientifically interesting and technically challenging. First we examine the high entropy alloy MoNbTaW~\cite{Senkov2010,WidomMCMD2013}. Here the chemically disordered structure makes the conventional phonon-based approach time consuming. Unfortunately it also increases the demands on AIMD run times and limits our ability to improve statistics through symmetrization. Next, we address the temperature-driven HCP to BCC transition of Ti. Owing to the presence of imaginary frequency modes in the BCC state, the usual harmonic and quasiharmonic approaches cannot be applied, while our method succeeds.

\section{Methods}
\label{sec:Methods}

\subsection{Probability density function}
\label{sec:PDF}
Our approach focuses on the $N$-body probability density function $\rho_N(\mathu,\mathp)$ of a classical $N$-atom system in Cartesian phase space. The displacement variable $\mathu=(\bu_1,\bu_2,...,\bu_{N})$, where $\bu_i\equiv \br_i-\bR_i$ defines atomic displacement of the position $\br_i$ of atom $i$ from its mean position (ideal lattice site) $\bR_i$, and $\mathp=(\bp_1,\bp_2,...,\bp_N)$ incorporates the momenta $\{\bp_i\}$. Owing to the additivity of kinetic and potential energy, the phase space probability factors into a product of density functions $f_u(\mathu)$ and $f_p(\mathp)$
\begin{equation}
  \label{eq:2}
  \rho_N(\mathu,\mathp)=h^{3N}f_u(\mathu)f_p(\mathp).
\end{equation}
The factor $h^{3N}$ comes from the constraint that the probability density integrates to 1,
\begin{equation}
  \label{eq:3}
  \frac{1}{h^{3N}}\iint\cdots\int_{-\infty}^\infty \rho_N(\mathu,\mathp)
  d^{3N}\,\mathu\,d^{3N}\mathp=1.
\end{equation}
The entropy according to Gibbs~\cite{Gibbs1902} is
\begin{equation}
  \label{eq:S:info}
  S=-\frac{1}{h^{3N}}\iint\cdots\int_{-\infty}^\infty \rho_N(\mathu,\mathp)\ln
  \rho_N(\mathu,\mathp)\,d^{3N}\mathu\,d^{3N}\mathp.
\end{equation}
This is identical to the Shannon~\cite{Shannon1948} information-theoretic entropy, in suitable units.

According to classical Maxwell-Boltzmann statistics, the momentum distribution function is Gaussian,
\begin{equation}
  \label{eq:5}
  f_\mathp(\mathp)=\frac{\exp(-\frac{1}{2}\mathp^{\rm T}\bm{\Sigma_p^{-1}}\mathp)}{\sqrt{(2\pi)^{3N}\det(\bm\Sigma_p)}},
\end{equation}
with $\Sigma_p$ a diagonal matrix of entries $m_i/\beta$ where $m_i$ is the mass of atom $i$ and $\beta=1/k_{\rm B}T$. Formally, we set $\bm{M}={\rm diag}\left(m_1,m_1,m_1,m_2 \cdots, m_N\right)$, so that $\Sigma_p=\bm{M}/\beta$.

In contrast to the simplicity of the momentum distribution, the density function $f_u(\mathu)$ is difficult to describe precisely, considering the many-body and anharmonic interactions among atoms. We choose to approximate it as a Gaussian with suitable covariance. Hence we write
\begin{equation}
  \label{eq:7}
  f(\mathu)=\frac{\exp(-\frac{1}{2}\mathu^{\rm T}\bm{\Sigma_u^{-1}}\mathu)}{\sqrt{(2\pi)^{3N}\det(\bm{\Sigma_u})}},
\end{equation}
where $\bm{\Sigma_u}$ is the covariance matrix
\begin{equation}
  \label{eq:8}
  \bm{\Sigma_u}=\left(
  \begin{array}{cccc}
    \bm{\sigma}_{1,1} & \bm{\sigma}_{1,2} & \cdots&\bm{\sigma}_{1,N}\\
    \bm{\sigma}_{2,1} & \bm{\sigma}_{2,2} & \cdots&\bm{\sigma}_{2,N}\\
    \vdots & \vdots & \ddots&\vdots\\
    \bm{\sigma}_{N,1} & \bm{\sigma}_{N,2} & \cdots&\bm{\sigma}_{,NN}\\
  \end{array}
\right).
\end{equation}
The $\bm{\sigma}_{i,j}$ element of $\bm{\Sigma_u}$ is the $3\times 3$ covariance matrix of the displacements $\bu_i$ and $\bu_j$ of the $i^{\rm th}$ and $j^{\rm th}$ atoms,
\begin{equation}
  \label{eq:9}
  \bm{\sigma}_{i,j}=\left(
  \begin{array}{ccc}
    \avg{x_i x_j} & \avg{x_i y_j} &\avg{x_i z_j}\\
    \avg{y_i x_j} & \avg{y_i y_j} &\avg{y_i z_j}\\
    \avg{z_i x_j} & \avg{z_i y_j} & \avg{z_i z_j}
  \end{array}
\right),
\end{equation}
with $x$, $y$, and $z$ the Cartesian coordinates of the displacement $\bu$. Diagonal elements of the covariance matrix yield the variances, {\em e.g.} for our cubic lattices $\bm{\sigma}_{i,i}=\avg{x^2}\bm{1}$. Due to the Gaussian approximation, the many-body density $f_u(\mathu)$ factors into a product of two-body correlations. Note that these two-body terms include anharmonic effects through the values of their covariances.

Within these approximations, the entropy $S$ of N atoms becomes
\begin{equation}
  \label{eq:11}
  S=\frac{1}{2}\ln(\det{(\bm{\Sigma_u})})+\frac{3}{2}\sum_{i=1}^N\ln(m_i/\beta\hbar^2)+3N.
\end{equation}
If all masses are equal, $S$ simplifies to
\begin{equation}
  \label{eq:S:cov}
  S=\frac{1}{2}\ln\left((2\pi e\Lambda)^{3N}\det{(\bm{\Sigma_u})}\right)
\end{equation}
where $\Lambda=\sqrt{2\pi\hbar^2/mk_B T}$ is the thermal de Broglie wavelength for mass $m$ at temperature $T$. Subject to the Gaussian approximation, our method resembles the approach of Morris and Ho~\cite{Morris1995}, who applied it to a one-dimensional model system. However, the formalism of Eq.~(\ref{eq:S:info}) applies generally, and we will examine corrections to the Gaussian approximation in Section~\ref{sec:anharmonic}.

\begin{figure*}[htpb]
  \centering
  \includegraphics[height=.25\textwidth]{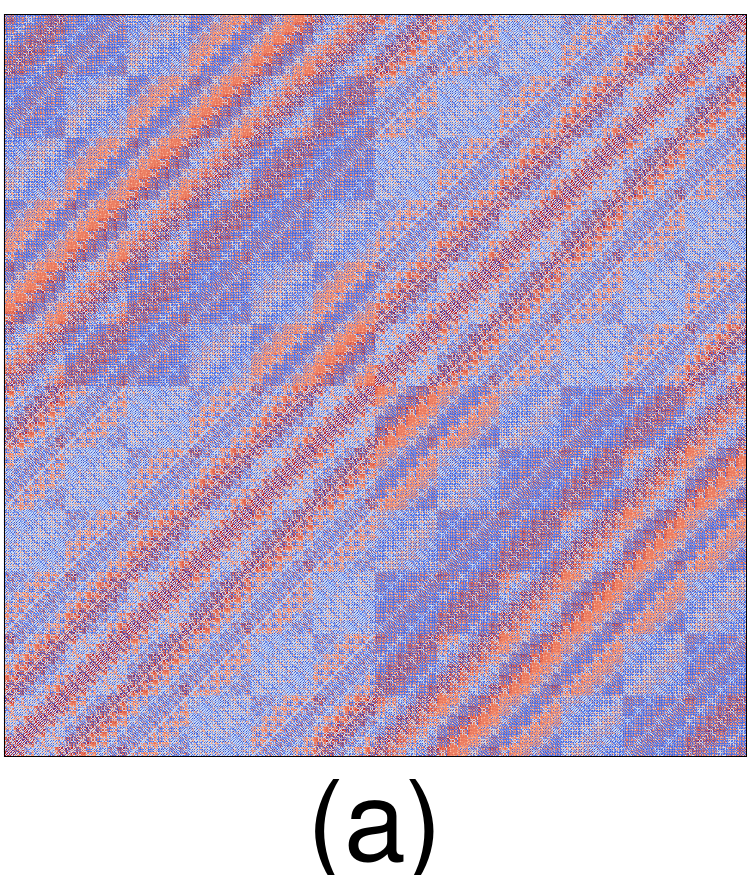}
  \includegraphics[height=.25\textwidth]{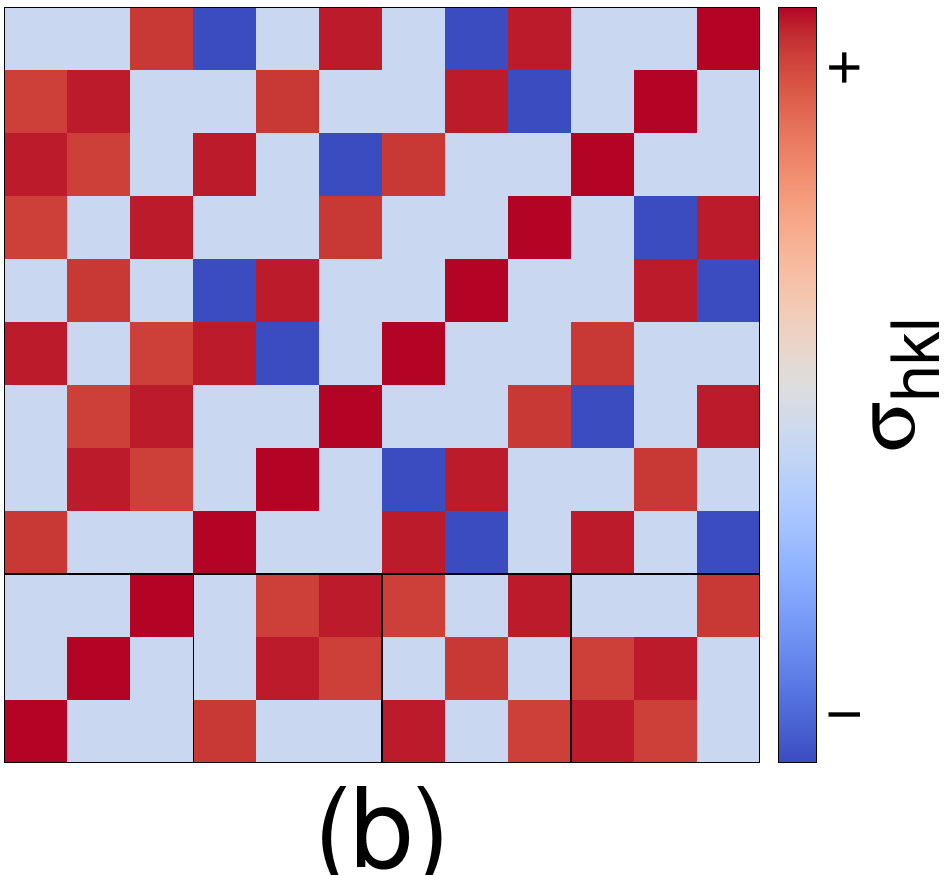}
  \includegraphics[height=.25\textwidth]{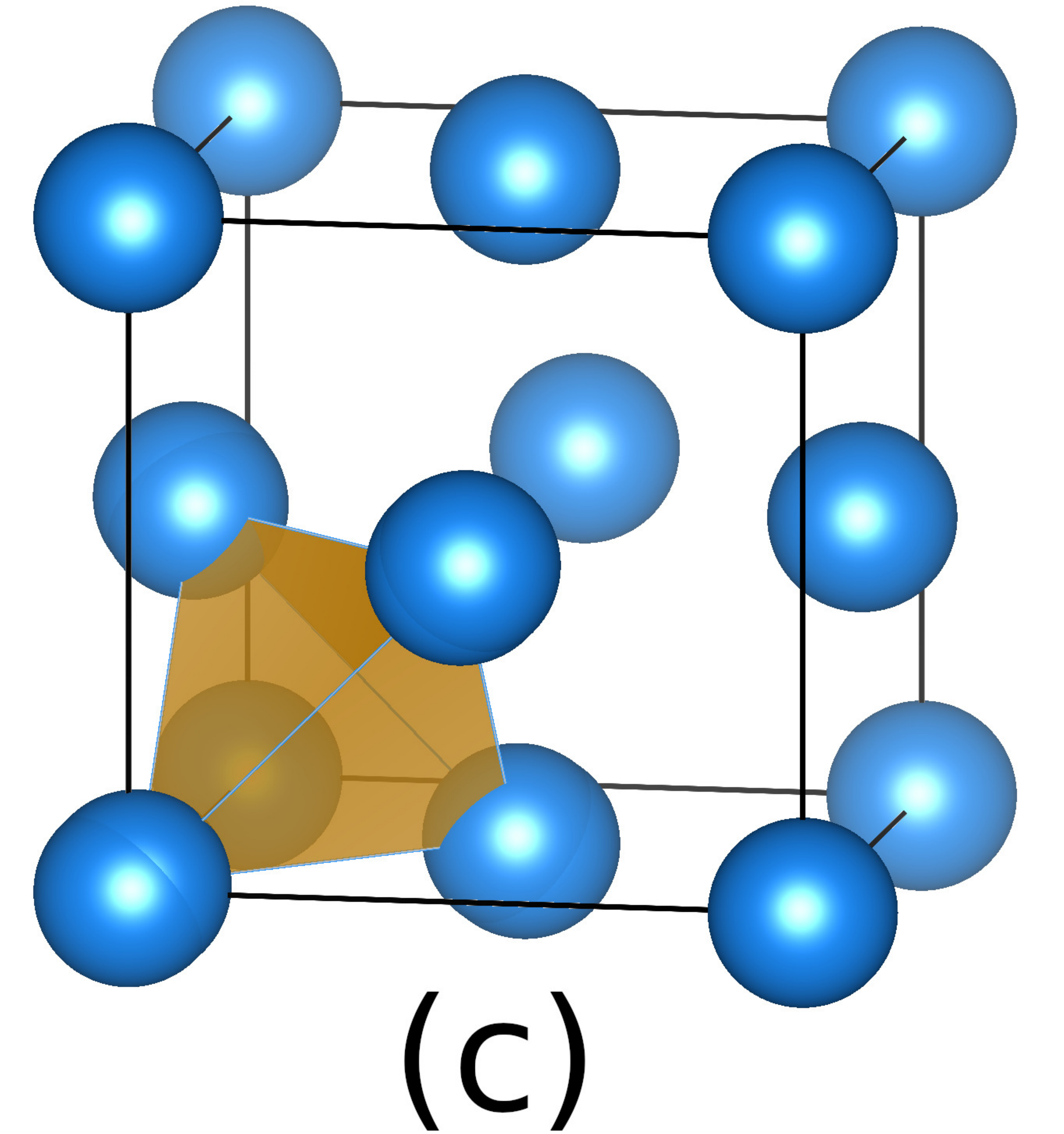}\\
  \includegraphics[width=.18\textwidth]{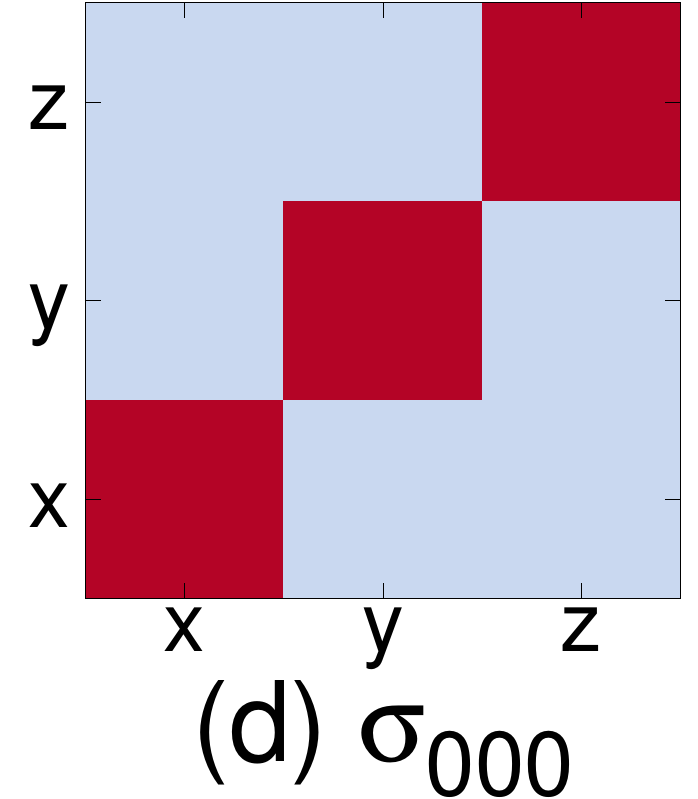}
  \includegraphics[width=.18\textwidth]{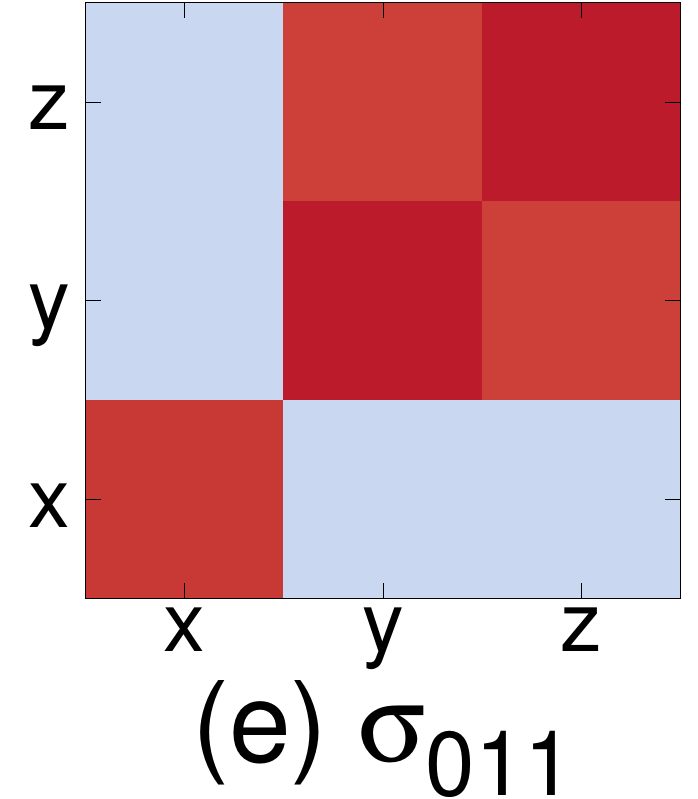}
  \includegraphics[width=.18\textwidth]{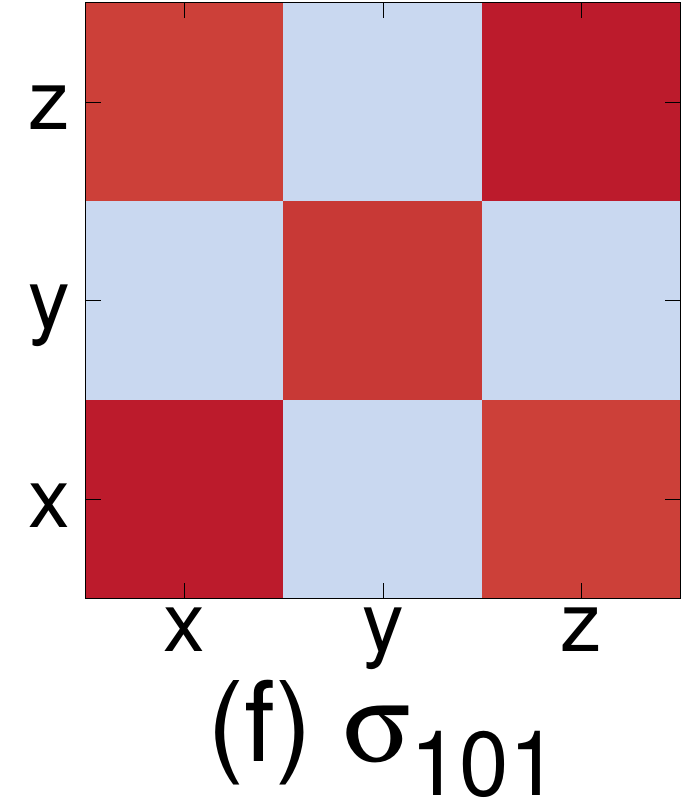}
  \includegraphics[width=.18\textwidth]{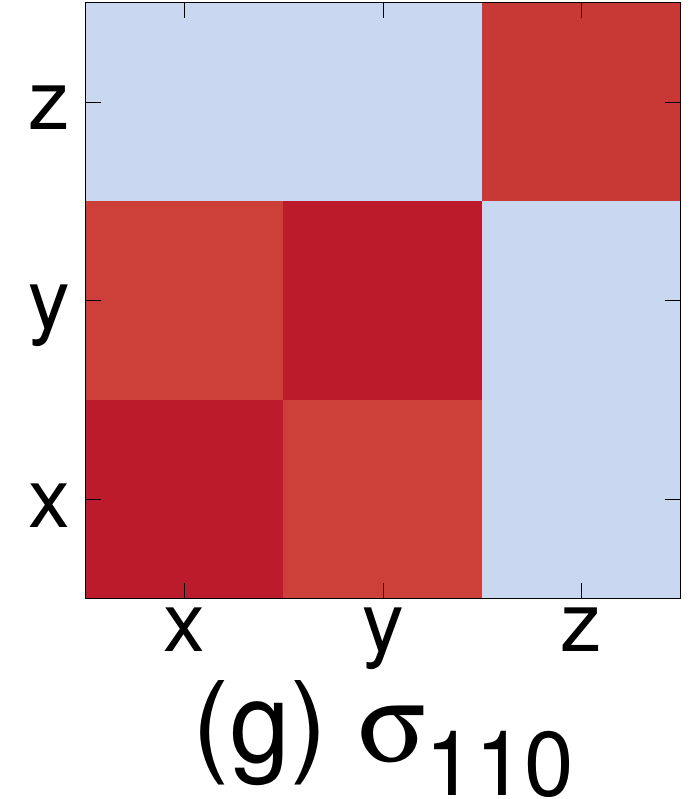}
  \caption{\label{fig:cm} (a) Complete AIMD simulated covariance matrix of FCC Al at 900K in a $4\times 4\times 4$ supercell of 256 atoms. (b) Submatrix of a 4-atom tetrahedron. (c) Unit cell of FCC Al illustrating tetrahedron of four nearest neighbors. (d)-(g) Single site variance matrix $\bm\sigma_{000}$ and three nearest-neighbor covariance matrices $\bm{\sigma}_{011}$, $\bm{\sigma}_{101}$ and $\bm{\sigma}_{110}$. Red color indicates positive covariance while blue color indicates negative covariance. Color bar indicates ${\rm sgn}(\sigma)\ln{(|\sigma/\sigma_{\rm min}|)}$.}
\end{figure*}

Fig.~\ref{fig:cm} illustrates the covariance matrix $\bm{\Sigma_u}$ for FCC Al at T=900K. Repeating patterns reflect the symmetries of the FCC structure. Translational symmetry requires that the covariance submatrix $\bm \sigma_{i,j}$ depends only on the relative position $\bm R_{i,j}=\bm R_j-\bm R_i=h\bm a+k\bm b+l\bm c$, of the $i^{th}$ and $j^{th}$ atoms. Consequently, covariance matrices $\bm \sigma_{i,j}$ sharing the same Miller indices $hkl$ share the same value, $\bm\sigma_{hkl}$. All $3\times 3$ matrices along the diagonal are equivalent and share the form  $\bm \sigma_{000}$ shown in part (d), whose off diagonal elements vanish due to mirror symmetries. Three-fold rotational symmetry can be seen in the covariance matrices $\bm\sigma_{011}$, $\bm\sigma_{101}$, $\bm\sigma_{110}$ (parts (e)-(g)) whose non-zero off-diagonal elements are $yz$, $xz$, and $xy$ components.

\subsection{Relation to force constant matrix}
\label{sec:forcematrix}
The probability density $\rho(x)$ of a classical oscillator in the
harmonic potential $U=\frac{1}{2}m\omega^2 x^2$ is
\begin{equation}
  \rho(x) = \sqrt{\frac{\beta m\omega^2}{2\pi}} e^{-\frac{1}{2}\beta m\omega^2 x^2},
\end{equation}
and the variance of its displacement is $\sigma^2=\left<x^2\right>=1/(\beta m\omega^2)$. The force constant $C=U''=m\omega^2$ is related to the variance by
$C=1/\beta\sigma^2$. For an $N$-particle system, the force constant matrix $C$ is defined in term of the second derivative of the potential $U$,
\begin{equation}
  C_{i\mu,j\nu}=\frac{\partial^2 U}{\partial u_{i\mu}\partial u_{j\nu}}.
\end{equation}
where $u_{i\mu},u_{j\nu}$ are elements of displacement $\mathu$ in which $i,j$ denote atoms and $\mu,\nu$ denote $x,y,z$ Cartesian coordinates. The mass-weighted covariance matrix, $\tilde \Sigma_{i\mu,j\nu}=\sqrt{m_im_j}\,\Sigma_{i\mu,j\nu}$, relates to the mass-reduced force constant matrix $\tilde C_{i\mu,j\nu}=C_{i\mu,j\nu}/\sqrt{m_im_j}$, by
\begin{equation}
  \label{eq:Inverse}
  \tilde C=\frac{1}{\beta}\tilde \Sigma^{-1},
\end{equation}
hence measurement of the covariance matrix yields the complete set of force constants. The matrices $\tilde C$ and $\tilde \Sigma$ are singular because of center of mass translation invariance. To invert the singular matrix, we represent $\tilde{\Sigma}=\sum_{k\mu}\lambda_{k\mu}\dya{k\mu}$ where $\{(\lambda_{k\mu}\equiv\beta \omega_{k\mu}^2)^{-1},\ket{k\mu}\}$ is the set of eigenvalues and eigenvectors of $\tilde{\Sigma}$. Then, noting that $\tilde{C}$ and $\tilde{\Sigma}$ share common eigenvectors, we invert the nonvanishing eigenvalues to obtain $\tilde{C}=\sum_{k\mu}\omega_{k\mu}^2\dya{k\mu}$.

For a harmonic potential $U$, the relationship Eq.~(\ref{eq:Inverse}) is exact; for an anharmonic system we may take Eq.~(\ref{eq:Inverse}) as defining temperature-dependent effective force constants.

\subsection{Quantum harmonic entropy}
\label{sec:quantum}

The entropies predicted by our classical theory agree quite well with the experimental values at high temperatures, but they fall below experiment at temperatures below the Debye temperatures $\Theta_D$, as seen in Fig.~\ref{fig:results}. The negative deviation is a consequence of the negative divergence of $\log{(u^2/\Lambda^2)}\sim 2\log{T}$ as $T\rightarrow 0$. Experimentally $S\rightarrow 0$ for all materials, by the third law of thermodynamics, because quantum mechanics inhibits the excitation of vibrational modes with frequencies greater than $\kB T/\hbar$.

To overcome the singularity of classical entropy, we adopt entropy of the quantum harmonic oscillator, using harmonic frequencies $\omega_{k\mu}$ obtained from eigenvalues of our covariance matrix as discussed in Sec.~\ref{sec:forcematrix}. Summing over the nonzero vibrational frequencies, the entropy with quantum corrections is
\begin{equation}
  \label{eq:quantumS}
  S=k_B\sum_{k\mu}\left[-\ln(1-e^{-\beta\hbar\omega_{k\mu}})+\frac{\beta\hbar\omega_{k\mu}}{e^{\beta\hbar\omega_{k\mu}}-1}\right].
\end{equation}
This yields better agreement when temperature is below the Debye temperature as shown in Fig.~\ref{fig:results}. In particular, the limit $S\rightarrow 0$ as $T\rightarrow 0$ is obeyed.

This quantum model is harmonic in the sense that it is exact for quadratic potentials $U$, but it incorporates anharmonicity through the effective vibrational frequencies which were derived from the simulated covariance matrix. Errors due to applying the quantum harmonic model should be small at low temperatures, where motion generically becomes harmonic. Some prior studies employ time-dependent velocity correlation functions, then Fourier transform over time to obtain frequencies~\cite{Rahman1964,Dickey1969,Lin2003}. The systematics of that approach differ markedly from ours, as in principle we do not require time evolution at all; we only simulate trajectories for the sake of enlarging our configurational ensemble.

The model Hamiltonian can be constructed in the actual harmonic limit of small oscillations by evaluating the force constants within density functional perturbation theory. This mode substantially underestimates the high temperature entropy as it neglects thermal expansion. The quasiharmonic approximation can be used to predict thermal expansion, resulting in improved agreement, or better yet we can evaluate the force constants at the experimental lattice parameters. As seen in Fig.~\ref{fig:results} the quasiharmonic approximation utilizing experimental lattice constants agree with experiment about as well as our new method.

\subsection{{\em Ab-initio} methods}
\label{sec:abinitio}

{\em Ab initio} molecular dynamics (AIMD) simulations are performed for FCC Al in supercells of size $4\times 4\times 4$ (256 atoms) and $6\times 6\times 6$ (864 atoms), and for BCC Na in a $6\times 6\times 6$ supercell (432 atoms). We use the Vienna Ab initio Simulation Package (VASP~\cite{Kresse96}) using augmented plane wave potentials~\cite{PAW} with the Perdew-Burke-Ernzerhof (PBE~\cite{PBE}) generalized gradient exchange correlation functional. We use a single electronic $k$-point and default plane wave energy cutoffs. When possible we use experimental lattice constants at the appropriate temperatures. The molecular dynamics simulations use Nos\'e thermostats with the default Nos\'e mass parameters. Our time steps are 2fs, and our runs extend to 40ps for Al ($4\times 4\times 4$) and 8ps for ($6\times 6\times 6$), and 7ps or greater for Na.

After allowing the simulated systems to approach equilibrium, the variances and covariances are calculated from a continuing simulation by averaging $\bu_i \bu_j$ over many samples. We also average over $\Omega$ reflection, rotation and translation symmetry operations ${\bf T}_k$ such that $\boldmath{\sigma}_{i,j}=\frac{1}{\Omega}\sum_k {\bf T}_k \bu_i \bu_j$ becomes symmetry invariant. In principle all the information needed to evaluate the entropy is contained in just a single representative structure of sufficient size, but the time averaging helps to reduce statistical error.

We perform phonon calculations as implemented in phonopy~\cite{phonopy} to obtain force constants and vibrational frequencies, and then calculate vibrational entropy as discussed in Section~\ref{sec:quantum}. Rather than calculating the thermal expansion {\em ab-initio}, as in the traditional quasiharmonic approximation~\cite{phonopy-qha}, we simply evaluate the force constants at the experimentally known temperature-dependent lattice constant $a(T)$.

Electronic entropy is evaluated as
\begin{eqnarray}
  \label{eq:Se}
  S=&-k_B\int {\rm d}E D(E) [f_{T,\mu}(E)\ln{f_{T,\mu}(E)} \nonumber\\
    &+(1-f_{T,\mu}(E))\ln{(1-f_{T,\mu}(E))}]
\end{eqnarray}
with $D(E)$ the electronic density of states calculated from a structure with lattice constant $a(T)$, and $f_{T,\mu}$ the Fermi-Dirac occupation function. The chemical potential $\mu$ is obtained as a function of $T$ using the program {\tt Felect}~\cite{Felect}.

\section{Applications}
\subsection{Test cases: FCC Al and BCC Na}
\label{sec:TestCases}

Our method successfully predicts vibrational entropy for Al and Na, as shown in Fig.~\ref{fig:results} parts~(a) and~(b). Fig.~\ref{fig:results} parts (c) and (d) compare the residual errors of various approximations by subtracting off the experimental entropies. Curves labeled ``Debye-Waller'' and ``single-site'' neglect correlations among the displacements of different atoms. In this case the entropy reduces to
\begin{equation}
  \label{eq:single-site}
  S=\frac{3}{2}\ln{\left[2\pi e(\sigma_x/\Lambda)^2\right]}
\end{equation}
where $\sigma_x^2=\langle u_x^2\rangle$ is the mean square displacement. This quantity is related to the Debye-Waller factor~\cite{Warren1969} that diminishes the diffraction intensity of a peak or wavevector $\bm{q}$ by the factor $\exp{(-q^2\langle u^2\rangle/3)}$. The displacements are sometimes given in terms of $B=8\pi^2\langle u^2\rangle/3$. In Fig.~\ref{fig:results} we compare the experimental entropies of Al and Na with the prediction of Eq.~(\ref{eq:single-site}) using experimental values of the B-factor. Given the seeming disparity between crystallographic and thermodynamic methods, the agreement is quite striking.

Note that the Debye-Waller and single-site entropies exceed the experimental values. The displacement of a single atom applies forces that displace nearby atoms, reducing the total amount of information needed to specify a given configuration $\mathu$. A similar effect is found in the entropy of liquids, where the information content of pair correlation functions reduces the entropy below the value for an ideal gas at the same overall density~\cite{GaoWidom2018,WidomGao2019,HuangGaoWidom2021}. The improvement upon including the full covariance matrix is evident in the curve labeled ``classical'', but as discussed previously it suffers an unavoidable $\ln{T}$ negative divergence. This divergence is aleviated at low temperatures through quantum model (Sec.~\ref{sec:quantum}). The quasiharmonc model is also quite accurate.

\begin{figure*}[htpb]
  \centering
  \includegraphics[height=.35\textwidth]{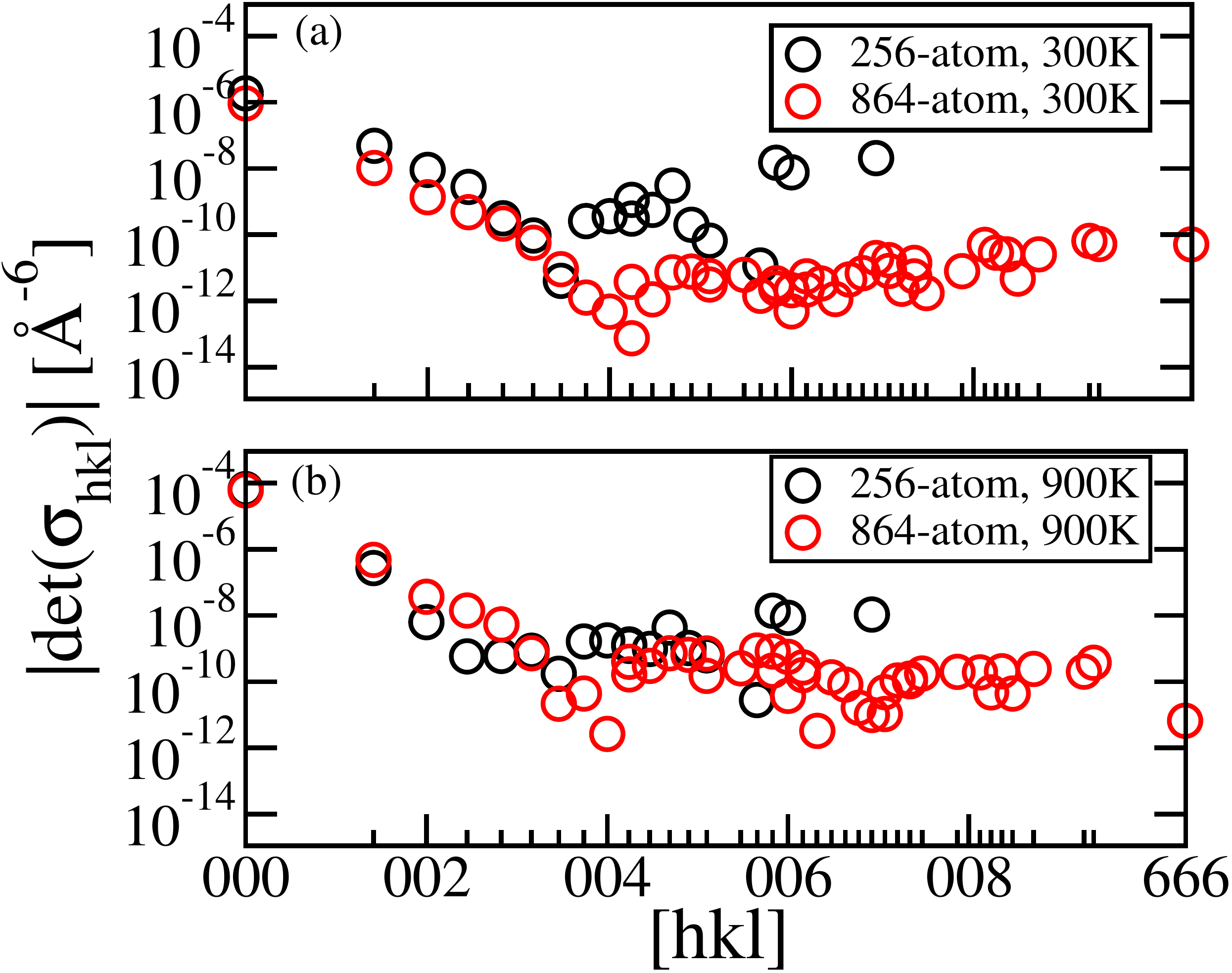}
  \includegraphics[height=.35\textwidth]{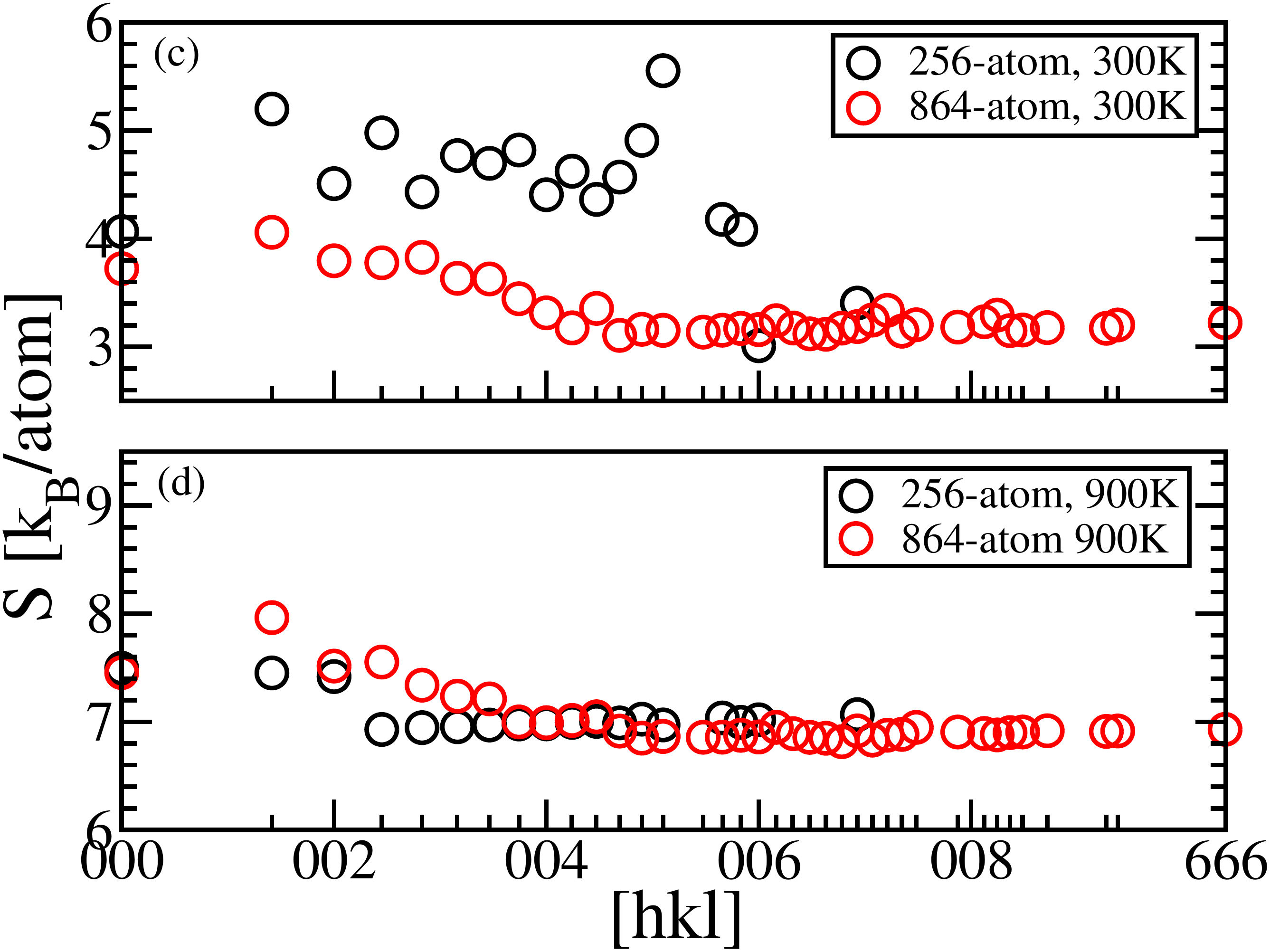}
  \caption{Left: Comparison of covariance matrix elements $\bm{\sigma}_{hkl}$ as bond length increases for Al at (a) $T=300$K and (b) $T=900$K. Right: Convergence of entropy $S$ after including all covariance matrix elements $\bm{\sigma}_{hkl}$ of pairs within $R<R_{hkl}$ at (c) T=300K and (d) $T=900$K. }
  \label{fig:size}
\end{figure*}

To better understand how the covariance matrix and entropy are influenced by the range of correlations, and by our finite MD simulation cells, we study the convergence of covariance matrix elements and corresponding entropy of Al, including only matrix elements $\bm{\sigma}_{hkl}$ of pairs separated by $R_{hkl}=|h\ba+k\bb+l\bc|$. Fig.~\ref{fig:size} (a) and (b) show that the absolute value of $\det{(\bm{\sigma}_{hkl})}$ drops rapidly with increasing the bond length, suggesting our simulation cell size is sufficient to capture the dominant collective motions of the solid, although some indication of cell size dependence can be seen in the excess correlation around $[hkl]=004$ at $T=300$K. Similar decay of correlations was observed in other simulations~\cite{Nicholson2021,Jeong2003}. Comparing $T=900$K with $T=300$K, we see similar variation with $R_{hkl}$, while the values at high temperature are nearly two orders of magnitude larger. Comparing convergence of the $4\times 4\times 4$ (256 atom) cell with the $6\times 6\times 6$ (864 atom) cell in Fig.~\ref{fig:size} (c) and (d) suggests the $4\times 4\times 4$ is fully adequate for entropy calculation at high temperatures, but just barely sufficient at low temperature.

\subsection{High entropy alloy: vibrational entropies of MoNbTaW}
Although high entropy alloys (HEAs) acquire their name from the entropy of chemical substitution, their vibrational entropy may exceed their substitutional entropy by a considerable margin. Substitutional entropy is relevant for stability mainly because the vibrational entropy of the mixture lies close to the average vibrational entropy of the elements~\cite{Gao2017}. Here, we investigate the applicability of our covariance method to calculate the vibrational entropy of MoNbTaW~\cite{Senkov2010}. Since chemical substitution is prevalent in HEAs, we have to choose what specific arrangement of atoms to take. We will take as representative structures the final configurations from hybrid MC/MD simulations~\cite{WidomMCMD2013}, which reflect the temperature variation of chemical order.

\begin{figure*}[htpb]
  \centering
  \includegraphics[width=.45\textwidth]{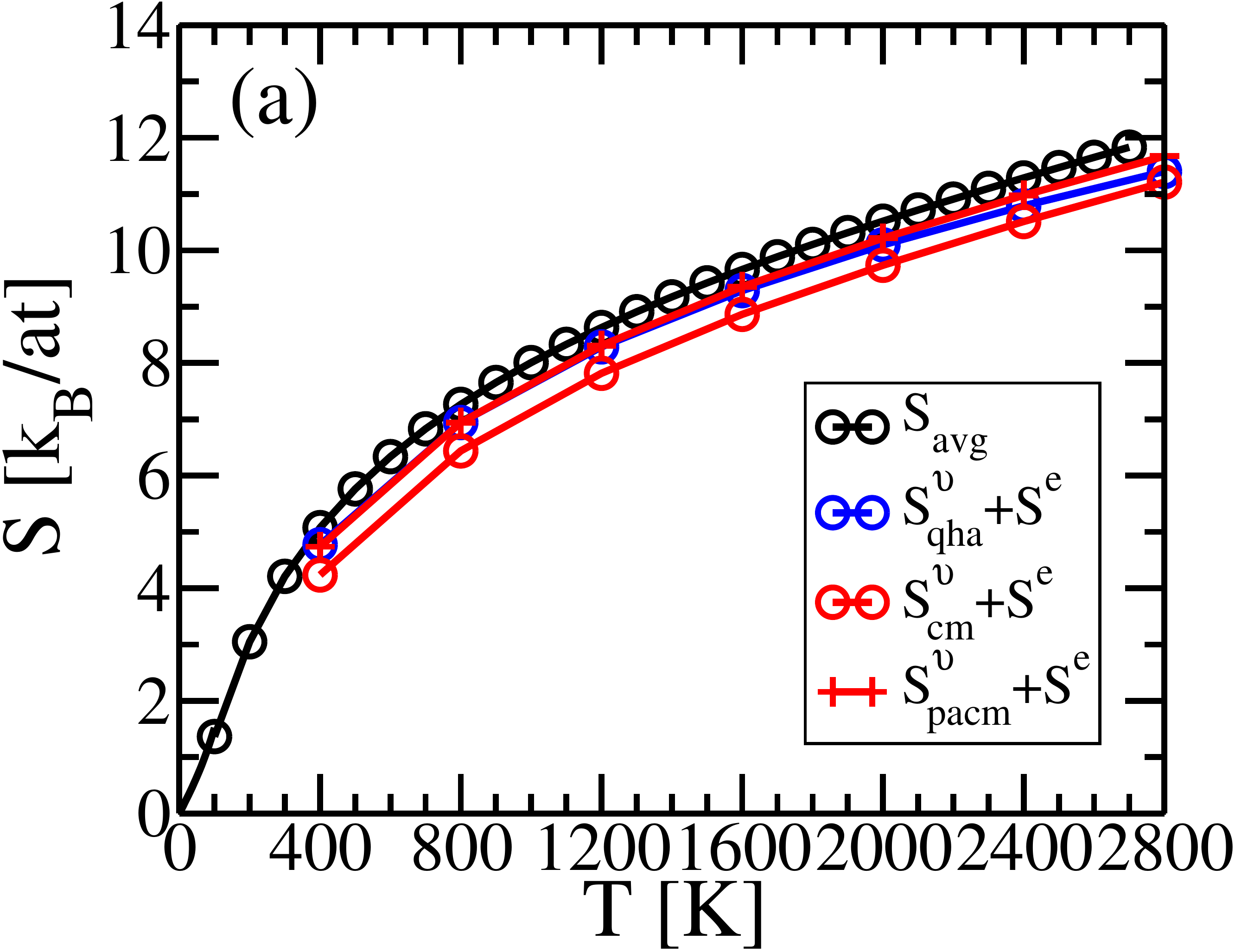}
  \includegraphics[width=.45\textwidth]{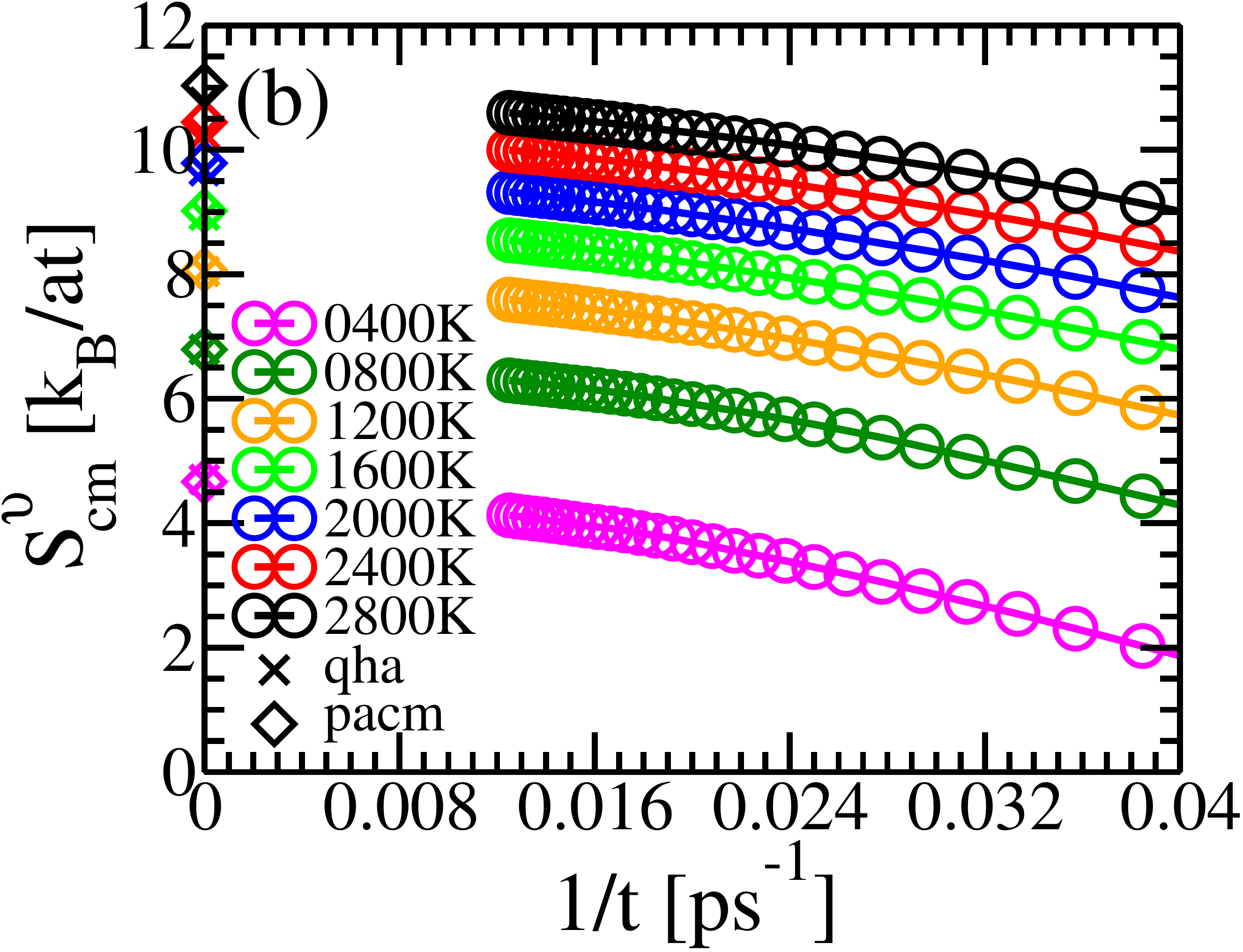}
  \caption{\label{fig:monbtaw}(a) Comparison plot of the average of experimental elemental entropies~\cite{JANAFMo,JANAFNb,JANAFTa,JANAFW}, $S_{\rm avg}$, with the entropy $S^v_{\rm qha}+S^e$ calculated using the quasiharmonic approximation, the entropy $S^v_{\rm cm}+S^e$ calculated from the covariance matrix, and the entropy $S^v_{\rm pacm}+S^e$ calculated from the pair-averaged covariance matrix. (b) Convergence of vibrational entropy $S^v_{\rm cm}$ as simulation time increases compared with vibrational entropies $S^v_{\rm qha}$ and $S^v_{\rm pacm}$.} 
\end{figure*}

We calculate the vibrational entropy $S^v_{\rm cm}$ of a specific chemical configuration at each temperature using the covariance matrix $\Sigma_\bu$ obtained from from MD simulations. Fig.~\ref{fig:monbtaw}~(a) plots entropies $S^v_{\rm cm}+S^e$ of theses structures. We compare our prediction with the average experimental entropies of pure elements, $S_{\rm avg}$, and with the quasiharmonic vibrational entropies $S_{\rm qha}^v+S^e$ of a cF16 (Heusler) MoNbTaW structure at the same lattice parameters as our MD simulations. These temperature-dependent lattice parameters were determined by varying the volume until the simulated total pressures vanish on average. It is seen from Fig.~\ref{fig:monbtaw}~(a) that both quasiharmonic and covariance matrix entropies are close to, but slightly smaller than, the averaged entropy $S_{\rm avg}$ of pure elements, consistent with prior calculations~\cite{Kormann2017}.

The vibrational entropy derived from the covariance matrix converges slowly because these chemically disordered structures lack symmetry and we cannot employ symmetry averaging as discussed in Section~\ref{sec:abinitio}. As a result, the covariance matrix has poor statistics and is hard to converge as illustrated in Fig.~\ref{fig:monbtaw}~(b). Unfortunately we lack an extrapolation formula for entropy {\em vs.} simulation time. At long times these entropies converge towards entropies calculated from the quasiharmonic approximation $S_{\rm qha}$.

In an effort to alleviate the poor statistics, we introduce a ``pair averaged covariance matrix'', $\bar{\Sigma}_\bu$, that maintains the chemical identities at each site while averaging of their chemical environments. The $(i,j)$ element of the full covariance matrix $\Sigma_\bu$ is the $3\times 3$ matrix $\bm{\sigma}_{i,j}^{\alpha\beta}$, where the superscripts remind us that the chemical species at site $i$ is $c(i)=\alpha$ and the chemical species at site $j$ is $c(j)=\beta$. Let the $\mathcal{P}_{i,j}^{\alpha\beta}$ be the set of all pairs $(i',j')$ such that $\bR_{i',j'}=\bR_{i,j}$ and $c(i')=\alpha$ and $c(j')=\beta$. We define the $(i,j)$ element of $\bar{\Sigma}_\bu$ as
\begin{equation}
  \label{eq:pacm}
  \bar{\Sigma}_\bu(i,j)=\bar{\bm{\sigma}}_{i,j}^{\alpha\beta}
  = \frac{1}{N_{i,j}^{\alpha\beta}}\sum_{i',j'}\bm{\sigma}_{i',j'}^{\alpha\beta}
\end{equation}
where the sum runs over the set $\mathcal{P}_{i,j}^{\alpha\beta}$ containing $N_{i,j}^{\alpha\beta}$ elements. The entropy computed from $\bar{\Sigma}_\bu$ is expected to provide a close upper bound on $S_{\rm cm}^v$.

\subsection{BCC to HCP phase transition in titanium}
\label{sec:BCC2HCP}

Certain elements and compounds are so strongly anharmonic that the entropy simply cannot be calculated within the harmonic or quasiharmonic approximation. Elements in columns 3 and 4 of the Periodic Table undergo diffusionless (Martensitic) phase transformations from BCC ($\beta$-phase) stable at high temperature to HCP ($\alpha$-phase) stable at low temperature. Harmonic analysis predicts their BCC states to be mechanically unstable at low temperature because they exhibit imaginary vibrational frequency modes. Eigenvectors of these modes describe the transformation pathway~\cite{BurgersW.G1934Otpo,Feng2018}. The instability prevents application of conventional harmonic or quasiharmonic calculations of the entropy. Our calculation method circumvents this difficulty because it does not require the calculation of vibrational frequencies.

These structural phase transitions are of practical importance, motivating considerable efforts to predict transition temperatures and understand their mechanisms~\cite{vdWalle2015,PhysRevB.95.064101,vdWalle2020,PhysRevB.80.104116,SangiovanniD.G2019Sdia,HellmanO2011Ldoa}. Proposed methods include phase space partitioning~\cite{vdWalle2015,PhysRevB.95.064101,vdWalle2020}, effective force constant averaging~\cite{HellmanO2011Ldoa}, and an ``augmented lattice'' model~\cite{PhysRevB.95.064101}. Predicted transition temperatures range from 1095K to 1114K, in general agreement with  in agreement with the experimental transition temperature $T_c=1166K$~\cite{JANAFTi001}. We apply our covariance matrix method to calculate vibrational entropy and  predict the transition temperature $T_c=1060$K.

\begin{figure*}[htpb]
  \centering
  \includegraphics[width=.23\textwidth]{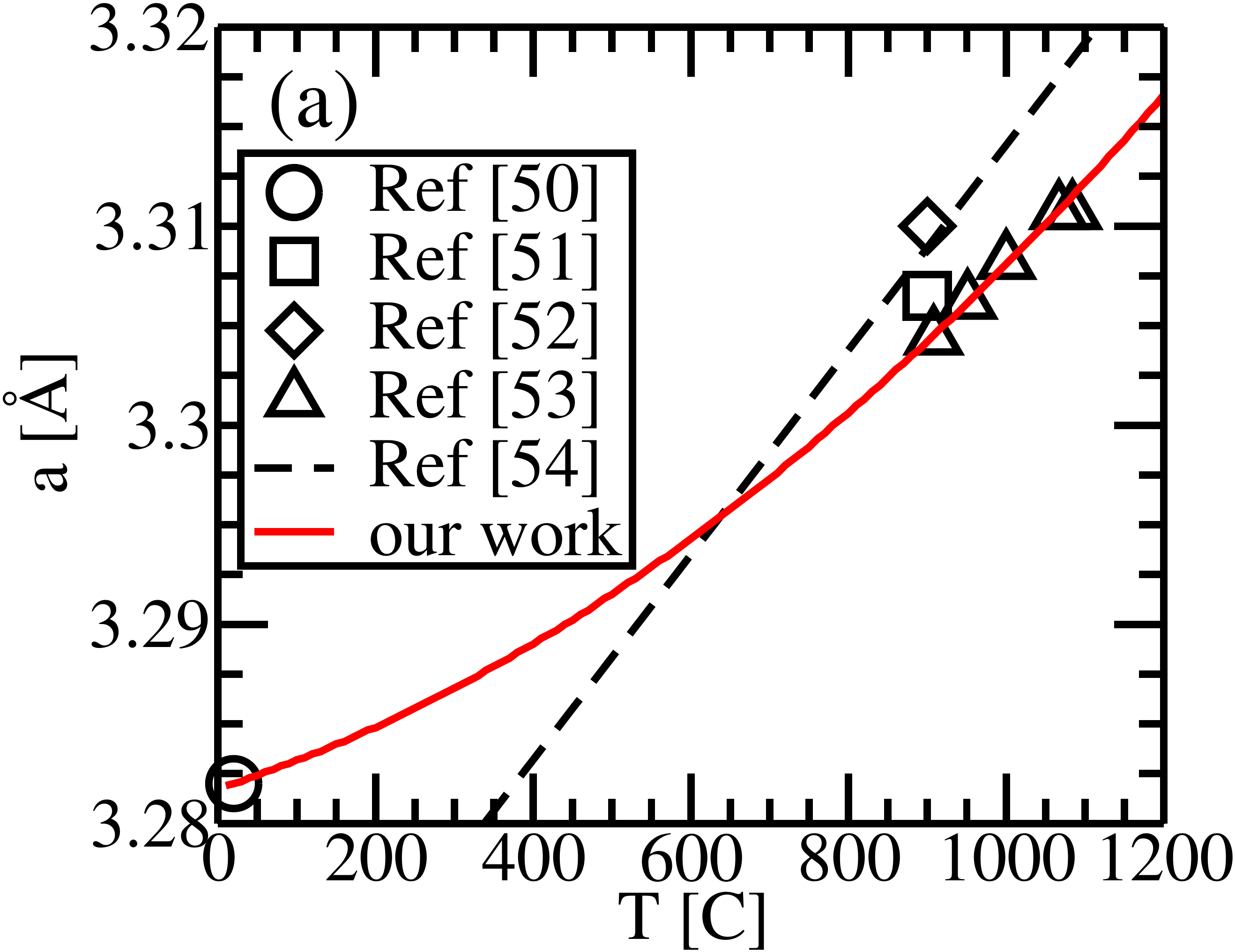}
  \includegraphics[width=.23\textwidth]{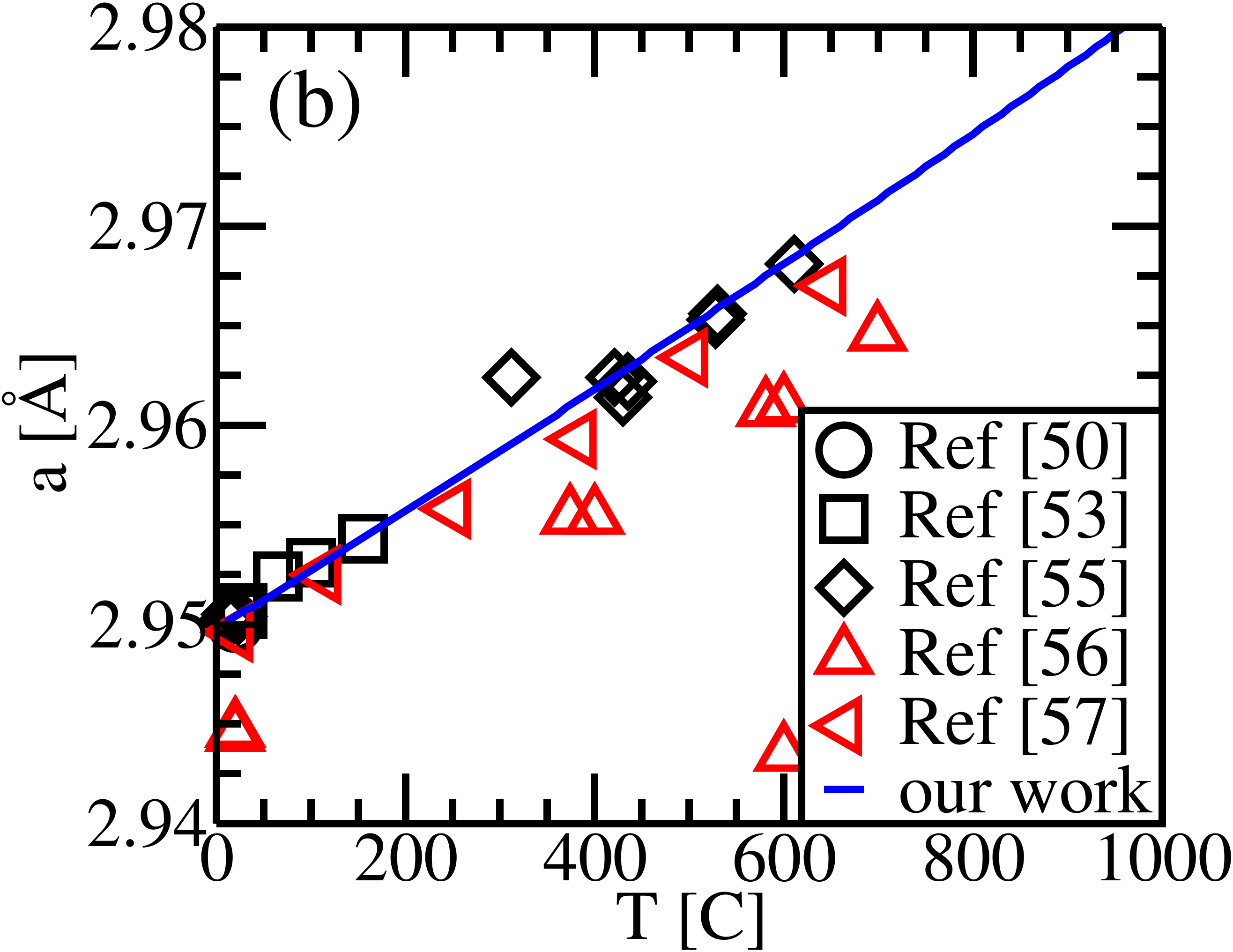}
  \includegraphics[width=.23\textwidth]{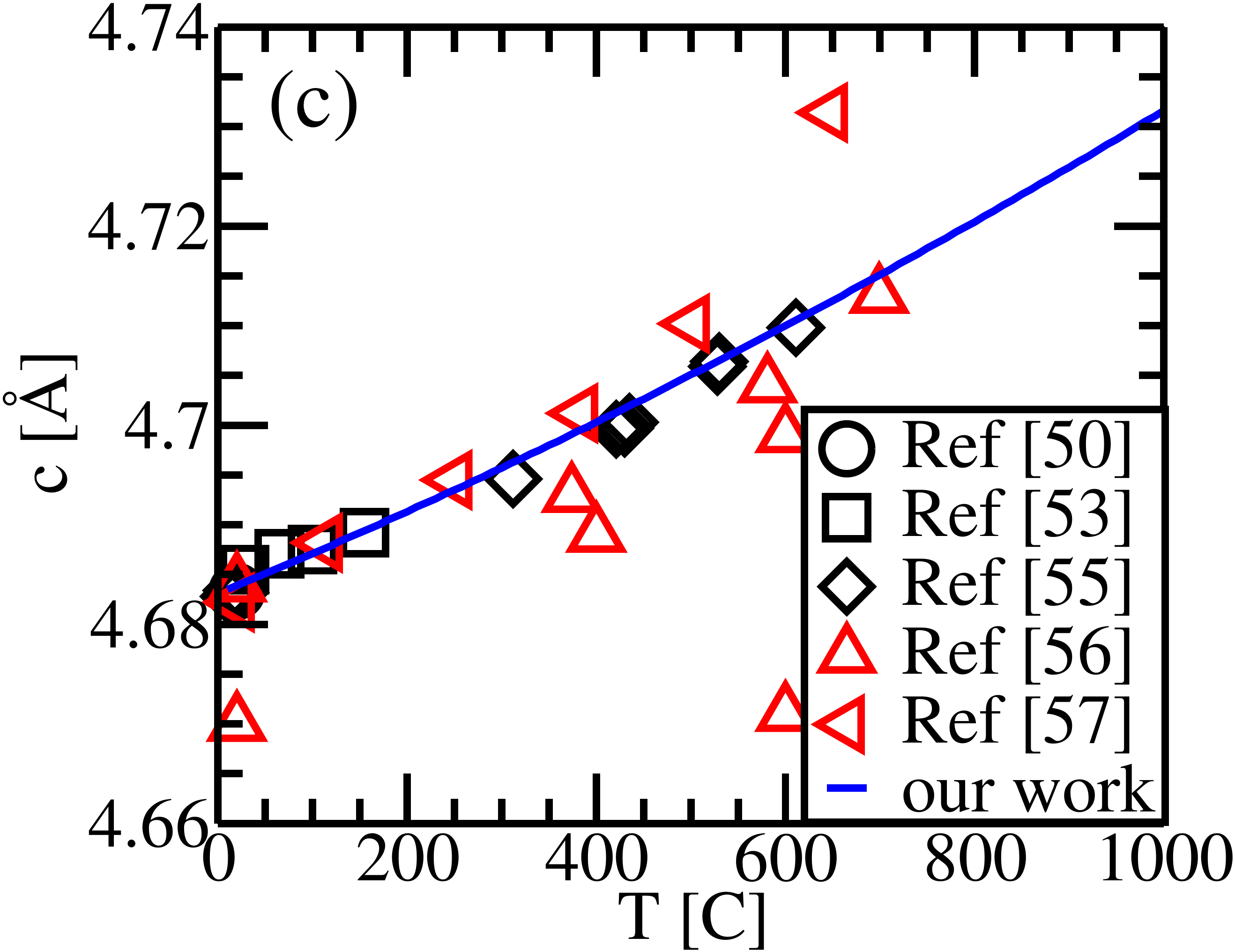}
  \includegraphics[width=.23\textwidth]{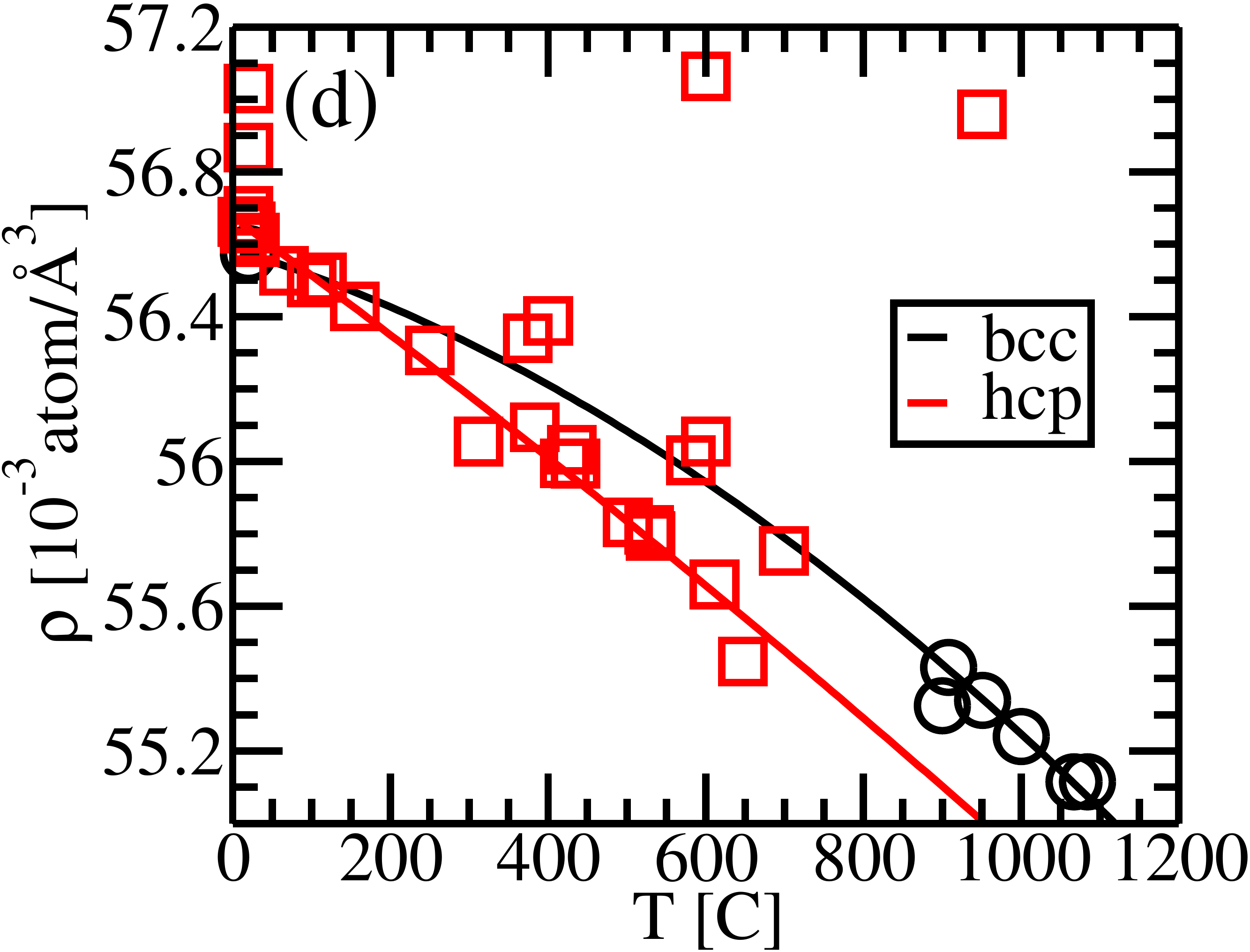}
  \caption{\label{fig:den} (a) Lattice constant $a$ of BCC Ti.
    \cite{AurelioG2002Mpit,EPPELSHEIMERDANIELS1950ADot,SenkovO.N2001Eota,SpreadboroughJ1959TMot,PETRYW1991Pdot}
    (b) and (c) Lattice constants $a$ and $c$ of HCP
    Ti.\cite{AurelioG2002Mpit,Pawar1968,RobertsW.T1962Poaa,SpreadboroughJ1959TMot,WillensR.H1962VXDf}
    Symbols in black are selected as fitting database. (d) Comparison of density $\rho$ between BCC and HCP Ti.}
\end{figure*}

We perform AIMD simulations for both BCC and HCP Ti at lattice constants that are fitted to experimental measurements with quadratic functions as shown in Fig.~\ref{fig:den}. Considering the scattering of experimental measurements of lattice constants, we choose to fit Ref.~\cite{AurelioG2002Mpit,SpreadboroughJ1959TMot} for lattice parameters
of BCC Ti and Ref.\cite{AurelioG2002Mpit,Pawar1968,RobertsW.T1962Poaa} for
HCP Ti. To minimize size effect, we prepare simulation cells with the same atomic number---an orthorhombic 256-atom 4x4x4 supercell based on a 4-atom unit cell $(\bm{a}=a,0,0;\bm{b}=0,a,-a;\bm{c}=0,a,a)$ for BCC Ti, and an orthorhombic 256-atom 4x4x4 supercell based on a 4-atom unit cell $(\bm{a}=a,0,0;\bm{b}=0,\sqrt{3}a,0;\bm{c}=0,0,c)$ for HCP Ti.

\begin{figure*}[htpb]
  \centering
  \includegraphics[width=.3\textwidth]{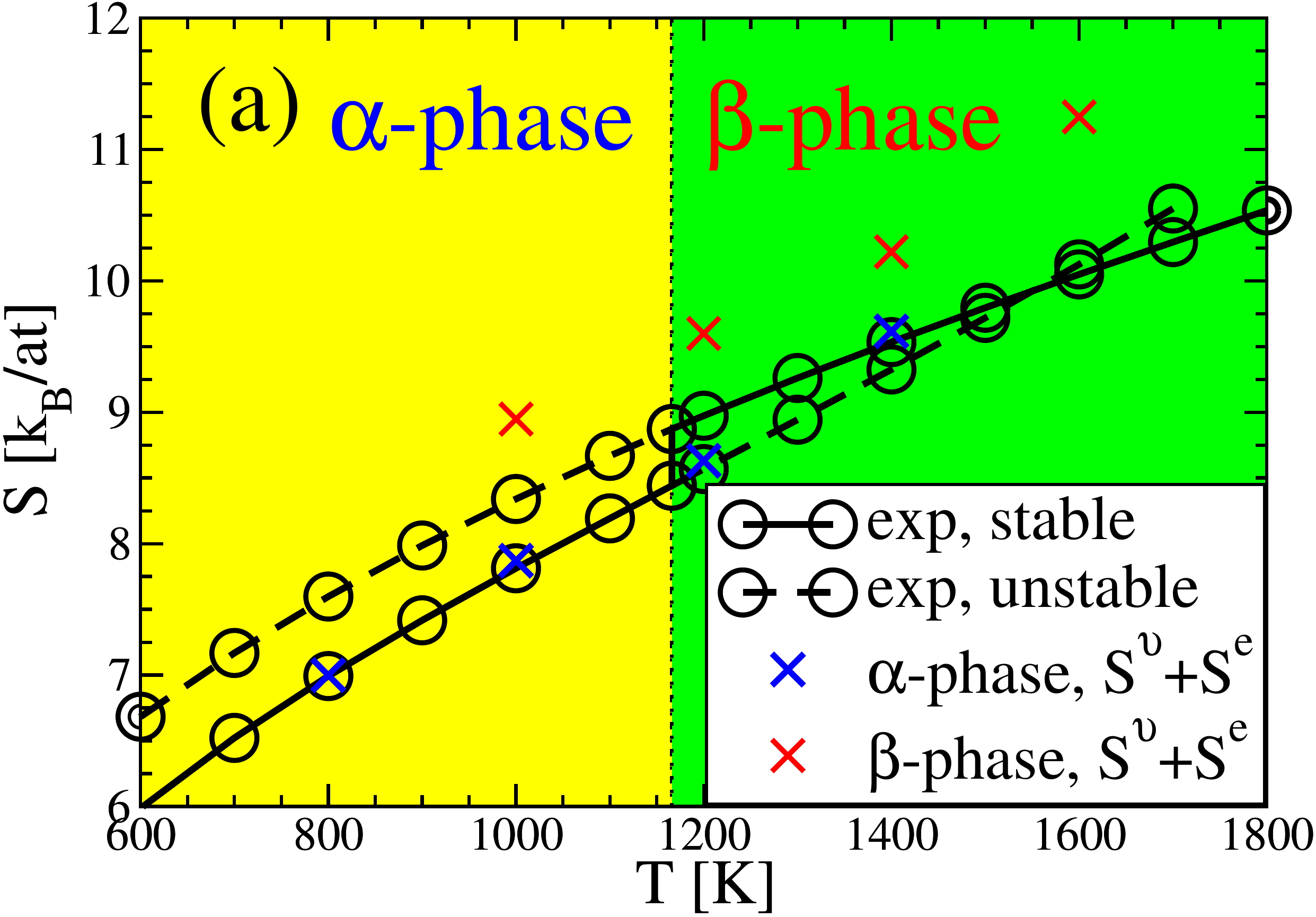}
  \includegraphics[width=.3\textwidth]{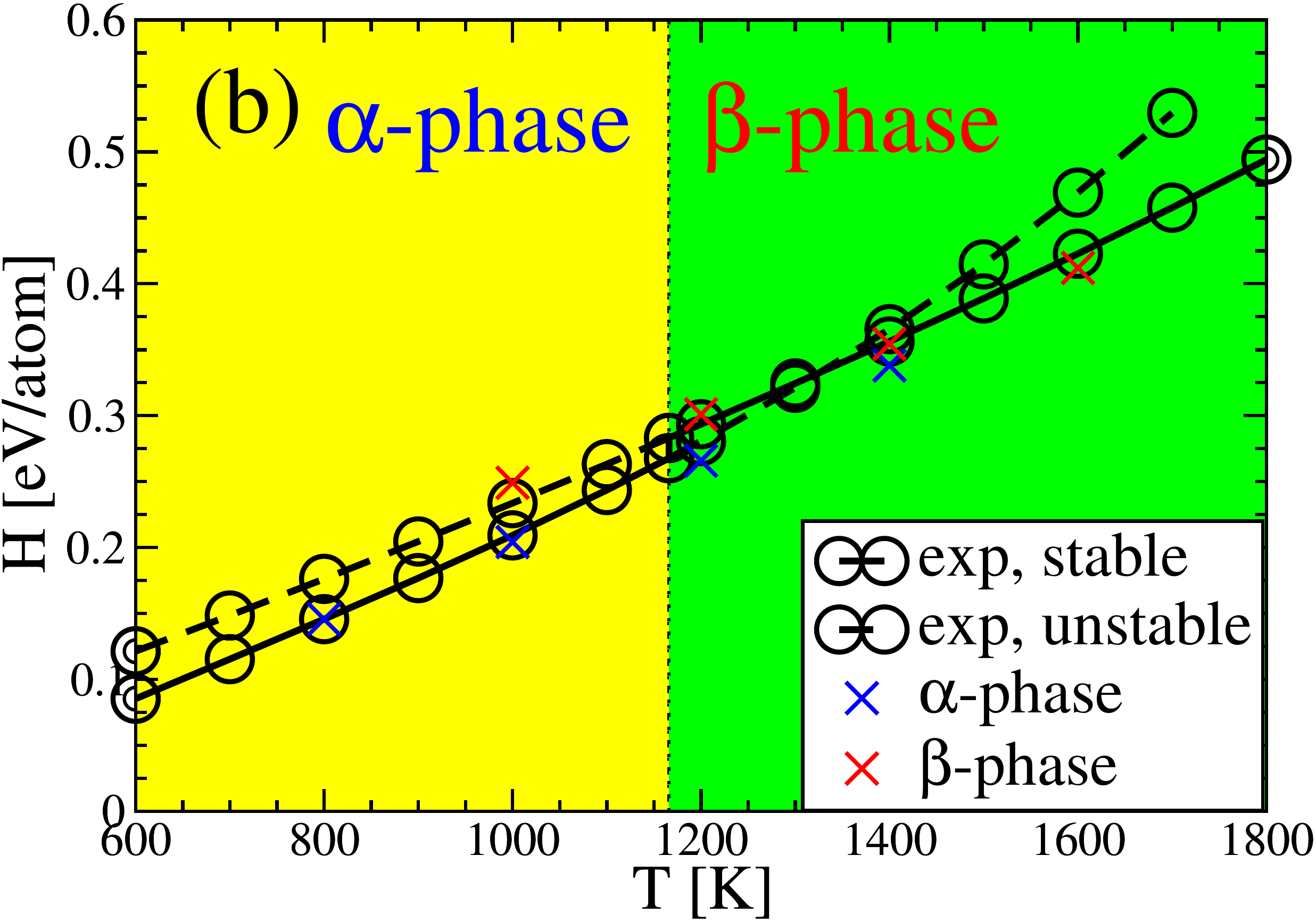}
  \includegraphics[width=.3\textwidth]{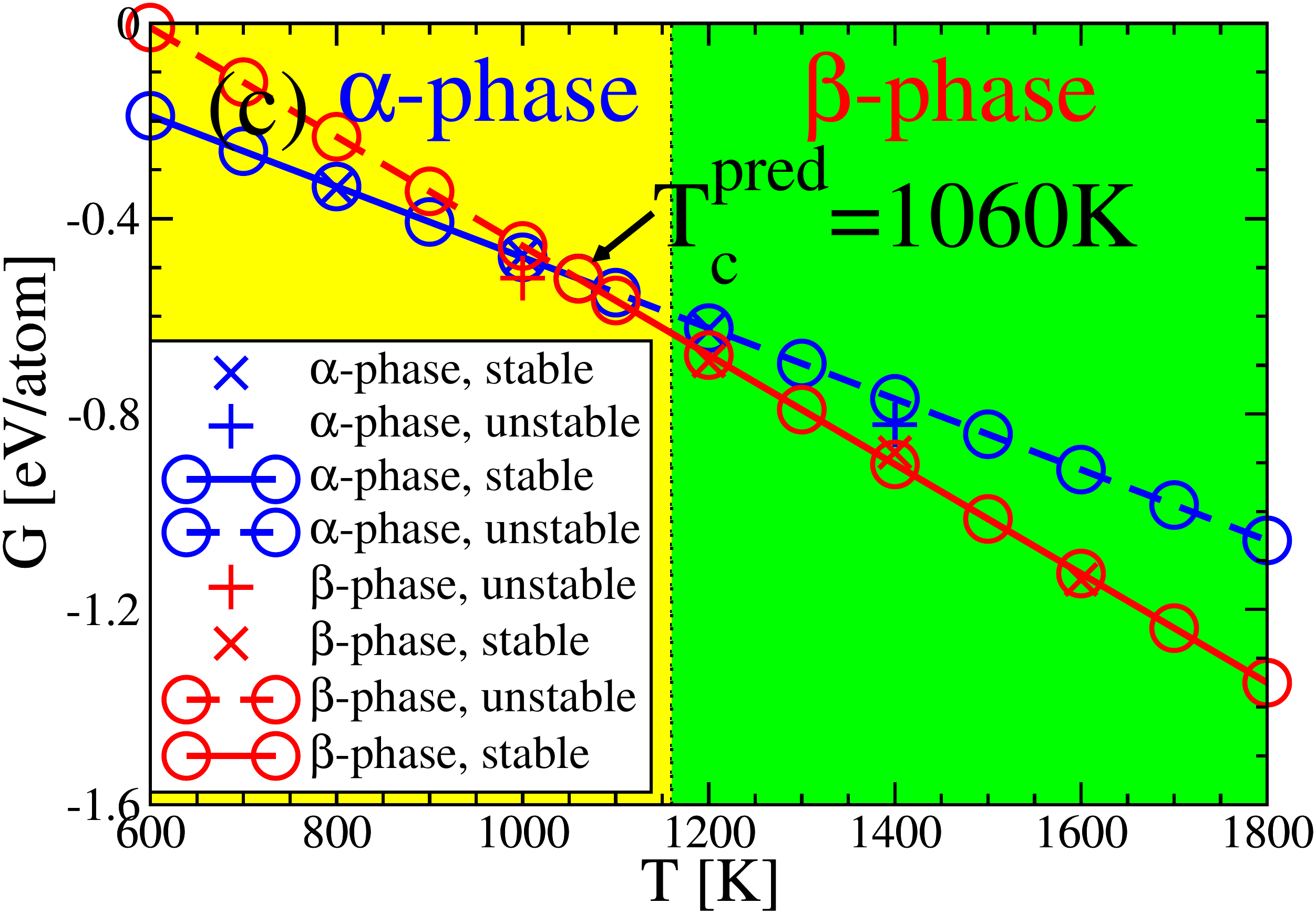}
\caption{\label{fig:ent} Comparison of calculated entropy and experimental
    entropy of $\alpha$-phase Ti and $\beta$-phase Ti. (b) Comparison of calculated and experimental enthalpy. (c) Comparison of experimental and calculated free energies. Yellow and green backgrounds shade regions of stability of the $\alpha$ and $\beta$ phases, respectively, as determined by experiment. Experimental data is plotted with solid lines in regions of stability, and dashed lines in regions of instability. Experimental data comes from NIST-JANAF Thermochemical Tables.\cite{JANAFTi001,JANAFTi002,JANAFTi003}. Calculated entropies $S=S^v+S^e$.}
\end{figure*}

\begin{figure*}[htpb]
  \centering
  \includegraphics[height=.38\textwidth]{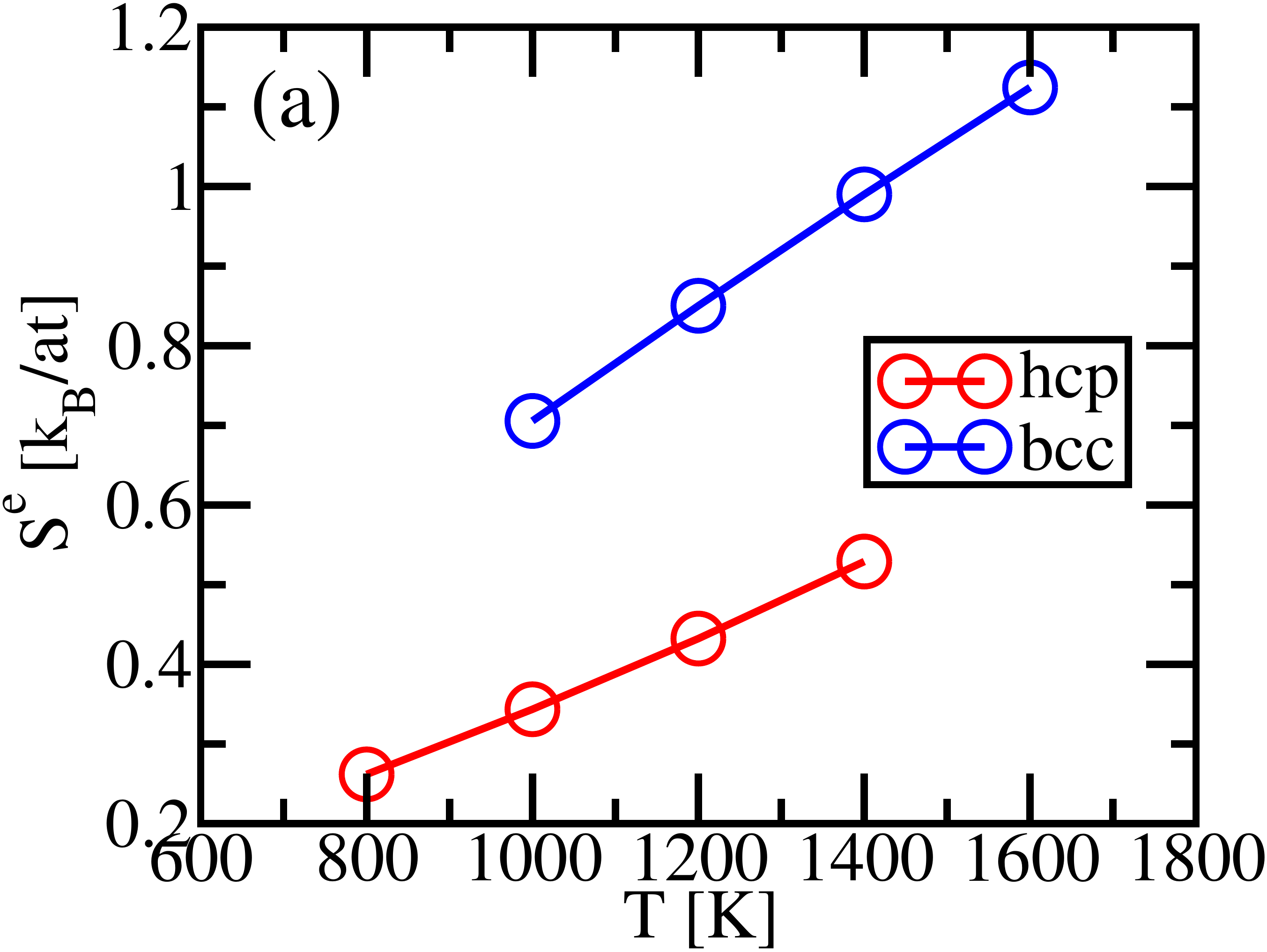}
  \includegraphics[height=.38\textwidth]{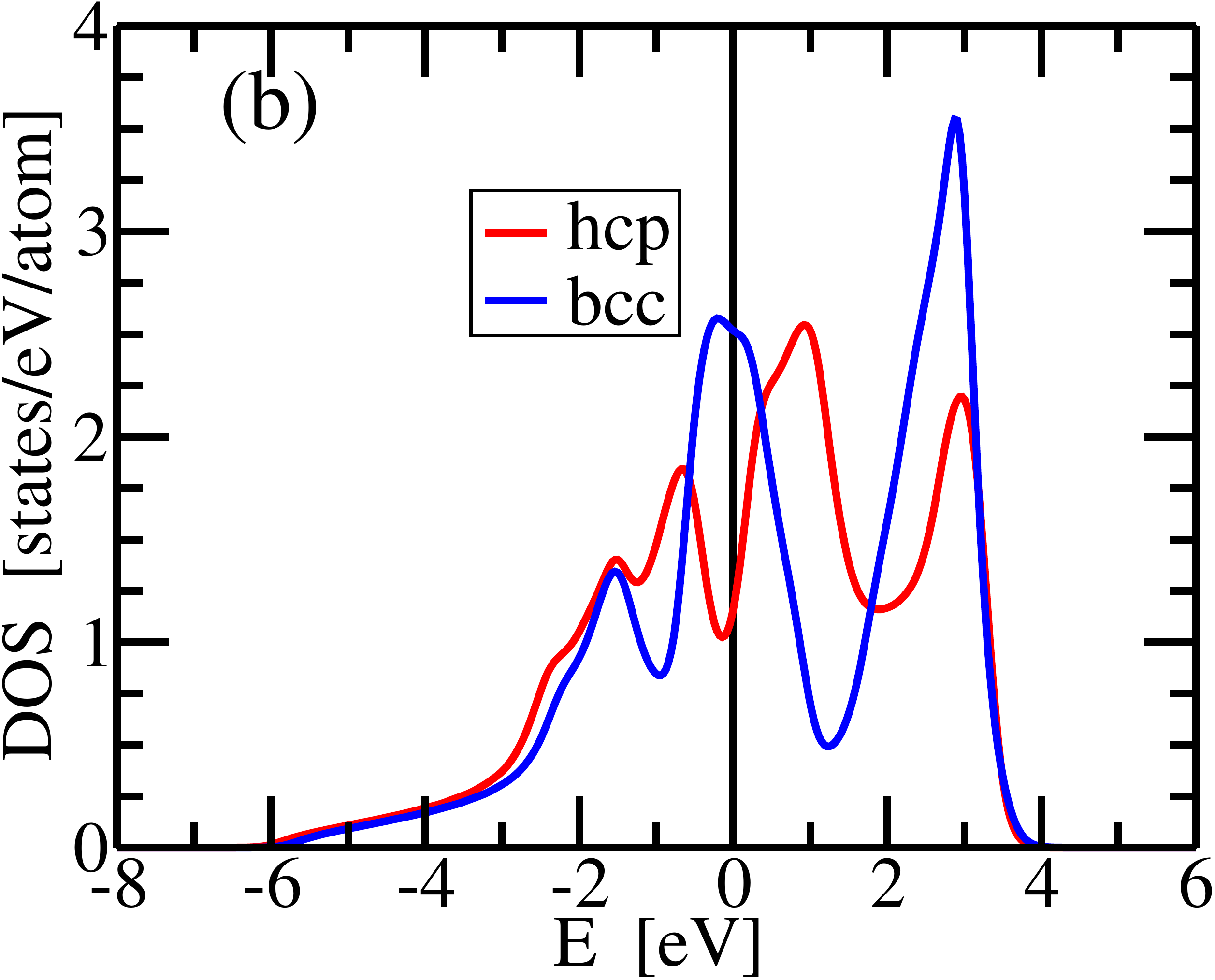}
  \caption{  \label{fig:DOS} (a) Calculated electronic entropies $S^e$ of BCC and HCP Ti. (b) Electronic densities of states of BCC and HCP Ti evaluated at their volumes at $T=1200$K. Fermi smearing of width $\sigma=\kB T=0.103$eV has been applied.}
\end{figure*}

A comparison of calculated total entropy $S^v_{cm}+S^e$ and experimental entropy is illustrated in Fig.~\ref{fig:ent}~(a). Electronic entropies $S^e$ are calculated from Eq.~\ref{eq:Se} with electronic density states obtained at the given volume for each temperature. As shown in Fig.~\ref{fig:DOS}, BCC Ti has a substantially higher electronic entropy than HCP Ti due to the pseudogap at the Fermi energy of the HCP density of electronic states. Formation of the pseudogap drives the  Burger's distortion from BCC to HCP~\cite{Feng2018}. Entropy of HCP Ti from our work compares well to the experimental entropy except one value at $T=1400$K which falls in the region where HCP is thermodynamically unstable. The entropy of BCC Ti, however, is overestimated by an amount of $0.5 k_{\rm B}$ to $1.0 k_{\rm B}$ at all temperatures.

Enthalpies are obtained by averaging energies over our MD simulations. To place enthalpies on the experimental scale, we shifted all of our calculated enthalpy values so that our enthalpy of $\alpha$ matched the experimental value at $T=800$K. For both phases our simulation matches well with measurement at temperatures below the $\alpha\to\beta$ temperature while at higher temperatures it falls below the experimental enthalpy.  Finally, we compute the Gibbs free energy $G=H-TS$ based on our calculated entropy and enthalpy of HCP Ti at T=$800$K, $1000$K, $1200$K and BCC Ti at T=$1200$K, $1400$K, $1600$K and predict $\alpha$-$\beta$ phase transition temperature $T_c^{\rm pred}=1060K$ (see Fig.~\ref{fig:ent}). 

\begin{figure*}[htpb]
  \centering
  \includegraphics[width=.4\textwidth]{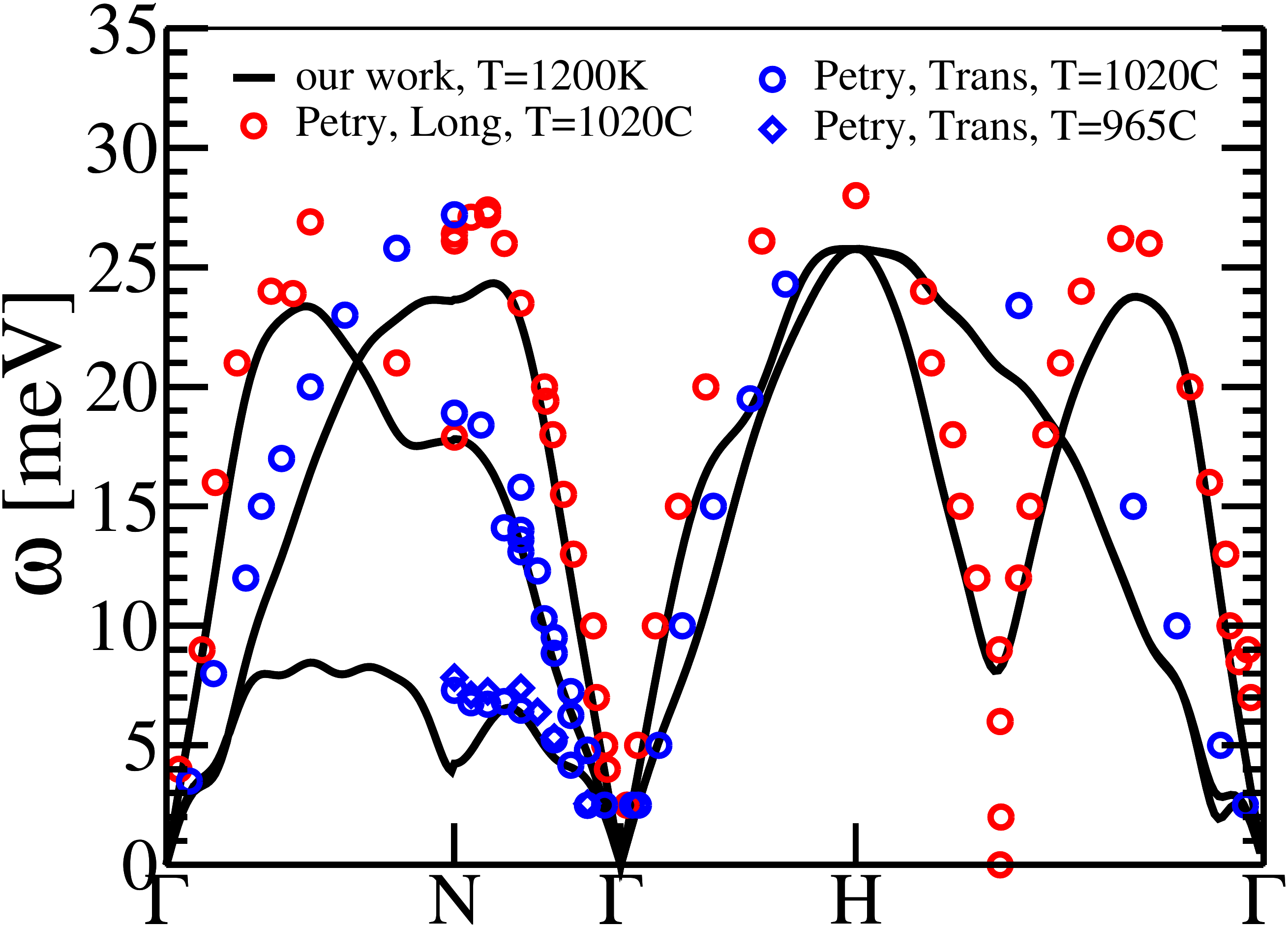}
  \includegraphics[width=.4\textwidth]{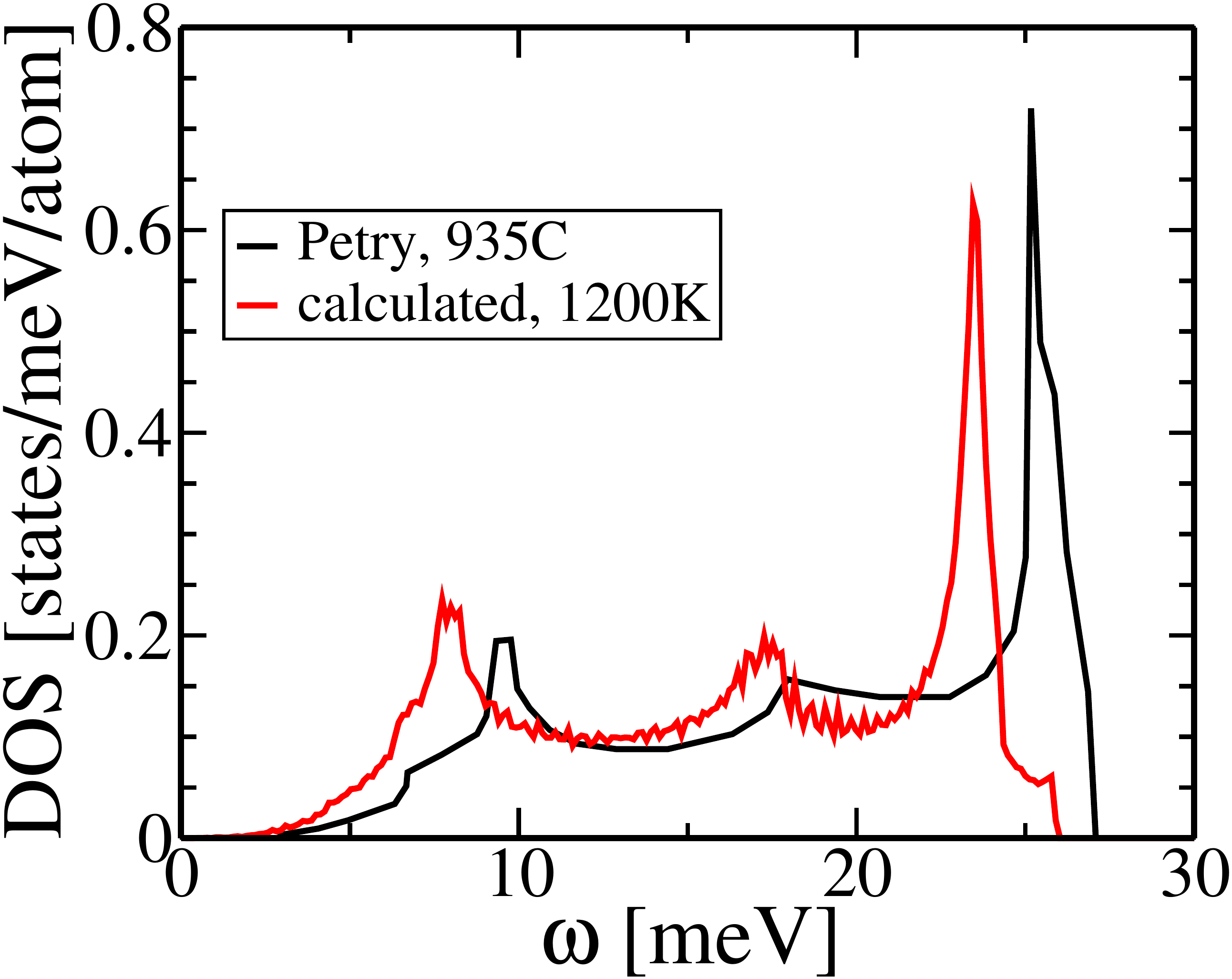}
  \caption{\label{fig:vband} (a) Calculated vibrational dispersion spectral of BCC Ti at
  T=1200K in a cubic cell with 250 atoms. (b) Calculated vibrational density of states. Red and blue dots in (a), and red curve in (b) come from Petry~\cite{PETRYW1991Pdot}.}
\end{figure*}

To understand the overestimate of BCC entropy, which leads to a low estimate of $T_c$, we compare calculated phonon spectra and vibrational density of states derived from our force constant matrix (see Fig.~\ref{fig:vband}) with results from Ref.~\cite{PETRYW1991Pdot}. Note that our effective vibrational frequencies fall systematically below the experiment, explaining the overestimate of entropy. We tested to see if this could be due to errors in lattice constant, but the impact of volume changes was not sufficient to explain our disagreement. Presumably the fault lies in some aspect of our simulation method. Below we investigate possible explanations in finite size effects, or anharmonicity, but these also turn out to be too small to explain the discrepancy.

\subsubsection{Finite site effect}
\label{sec:finite}
\begin{figure}[hptb]
  \centering
  \includegraphics[width=.4\textwidth]{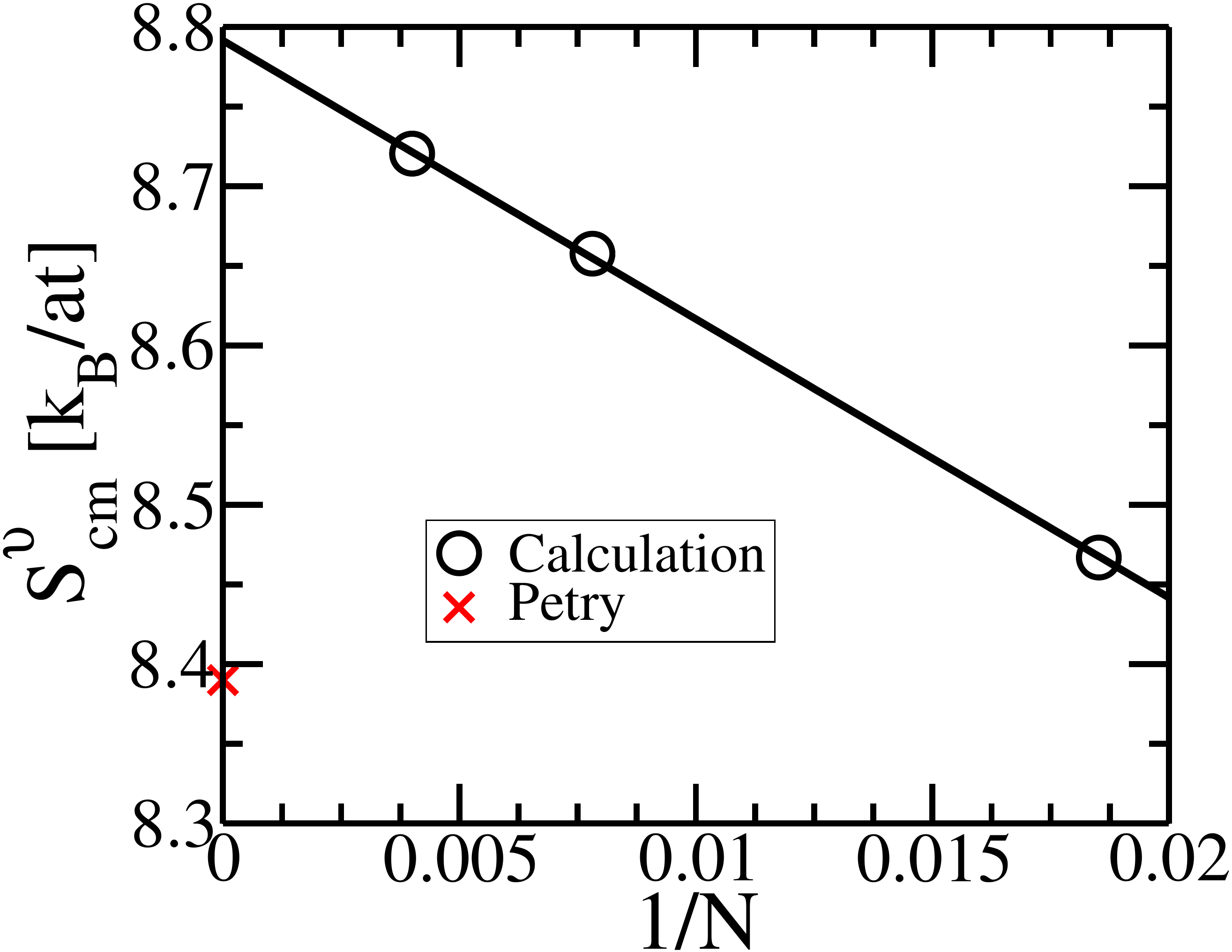}
  \caption{\label{fig:S_vs_N} Calculated vibrational entropy $S^v_{cm}$ versus the inverse of the simulated number of atoms, $1/N$, for BCC Ti at $T=1200$K. Experiment~\cite{PETRYW1991Pdot} is at $T=1208$K.}
\end{figure}

To evaluate the impact of simulated cell size on the entropy of BCC Ti, we perform entropy calculation for three sizes: 54-atoms, 128-atoms, and 250-atoms. Fig.~\ref{fig:S_vs_N} shows a linear relation between entropy $S^v_{cm}$ and inverse size $1/N$. With larger cells, entropy increases, and so does the disagreement with experiment. This finite size effect for BCC Ti resembles the finite size effect in high-pressure high-temperature BCC Fe~\cite{BelonoshkoAnatolyB2017Sobc}, so we believe the effect is real.

\subsubsection{Anharmonicity}
\label{sec:anharmonic}
We investigate the effect of anharmonic corrections to the single site probability density
\begin{equation}
  \label{eq:ph}
  p^h(\bu)=\frac{e^{-\frac{x^2+y^2+z^2}{2\sigma^2}}}{\sqrt{(2\pi\sigma^2)^3}}=\frac{e^{-\frac{R^2}{2\sigma^2}}}{\sqrt{(2\pi\sigma^2)^3}}.
\end{equation}
At the lowest order of anharmonicity, the probability density $p^a(\bu)$
includes the isotropic term
\begin{equation}
  \label{eq:anh}
  I(\bu)=(x^2+y^2+z^2)^2,
\end{equation}
and the anisotropic term
\begin{equation}
  \label{eq:ani}
  A(\bu)={{x}^{4}}+{{y}^{4}}+{{z}^{4}} -3 \left( {{x}^{2}} {{y}^{2}}+{{x}^{2}} {{z}^{2}}+{{y}^{2}} {{z}^{2}}\right),
\end{equation}
which are  invariant under cubic symmetry operations. The anharmonic probability density is hence approximated by
\begin{equation}
  \label{eq:pa}
  p^a(\bu)=\frac{1}{Z}\exp\left(-\frac{R^2}{2\sigma^2}\right)\exp\left(-\frac{a}{4\sigma^4}I(\bu)-\frac{b}{4\sigma^4}A(\bu)\right),
\end{equation}
where $Z$ is the normalization factor
\begin{equation}
  Z=\int_V\,d\bu~ p^a(\bu),
\end{equation}
and the integration volume $V$ is the Wigner-Seitz cell of an atom.

In practice we cut off the integration at the cube $V=[-8\sigma,+8\sigma]^3$, as justified by the rapid vanishing of $p^a(\bu)$. We calculate averages $\avg{R^2}$, $\avg{R^4}$, and $\avg{A}$ during our simulation, then we fit values of $\sigma$, $a$, and $b$ by solving the simultaneous nonlinear equations
\begin{eqnarray}
  \label{eq:simultaneous}
  \avg{R^2} &=&\int_V\,d\bu ~R^2~ p^a(\bu)\\
  \avg{R^4} &=&\int_V\,d\bu ~R^4~ p^a(\bu)\\
  \avg{A}   &=&\int_V\,d\bu ~A(\bu)~ p^a(\bu)
\end{eqnarray}
where the probability $p^a(\bu)$ is given by Eq.~(\ref{eq:pa}. Finally, the positional part of the anisotropic entropy, $S^a$, is calculated from
\begin{equation}
  S^a=-k_{\rm B}\int_V\,d\bu~ p^a(\bm u)\ln p^a(\bm u).
\end{equation}

\begin{table*}[htpb]
  \centering
  \begin{tabular}{|l|l|l|l||l|l|l|l|l|}
    \hline
    & $\left <R^2\right>$ [\AA] & $\left <R^4\right>$ [\AA$^4$] & $\left <I\right>$ [\AA$^4$] 
    & $\sigma^2$ [\AA$^2$] & $a$ & $b$ & $S^a-S^h$ [$k_{\rm B}$] \\
    \hline
    FCC Al & 0.05933&0.005945&\ 0.00003873 
    &0.01918&-0.005718&-0.001890& -0.00017975\\
    BCC Ti & 0.19061&0.06040&-0.002597 
    &0.06433&\ 0.003210&\ 0.01596& -0.001227\\
    \hline
  \end{tabular}
  \caption{\label{tab:solutions} Statistical average of $\left<R\right>$, $\left<R^2\right>$ from MD simulations. Correction to harmonic entropy and parameters  $\sigma^2$, $a$, $b$, and $S^a-S^h$ of FCC Al at $T=500$K and BCC Ti at $T=1200$K.}
\end{table*}

\begin{figure*}[htpb]
  \centering
  \includegraphics[width=.38\textwidth]{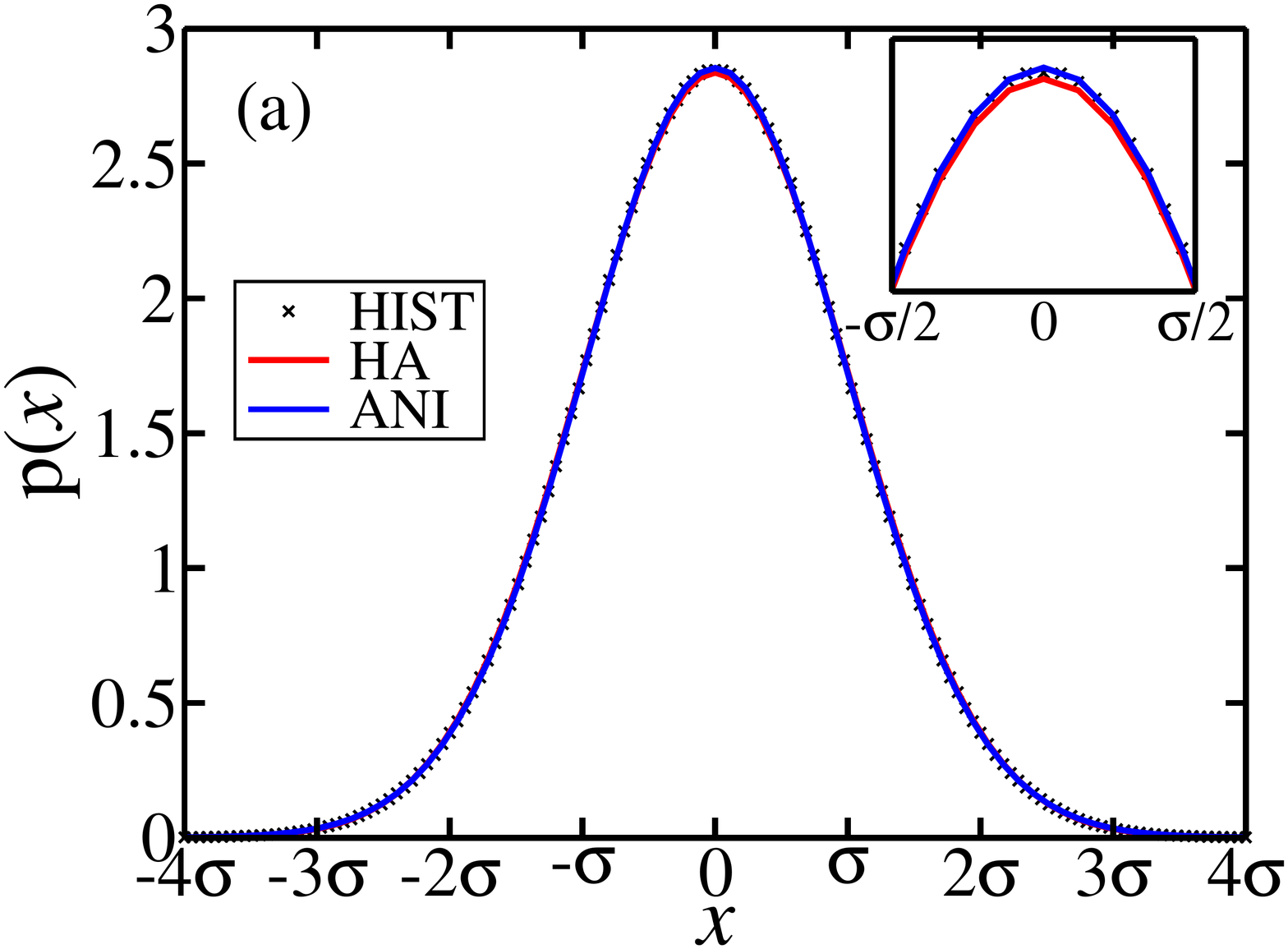}
  \includegraphics[width=.38\textwidth]{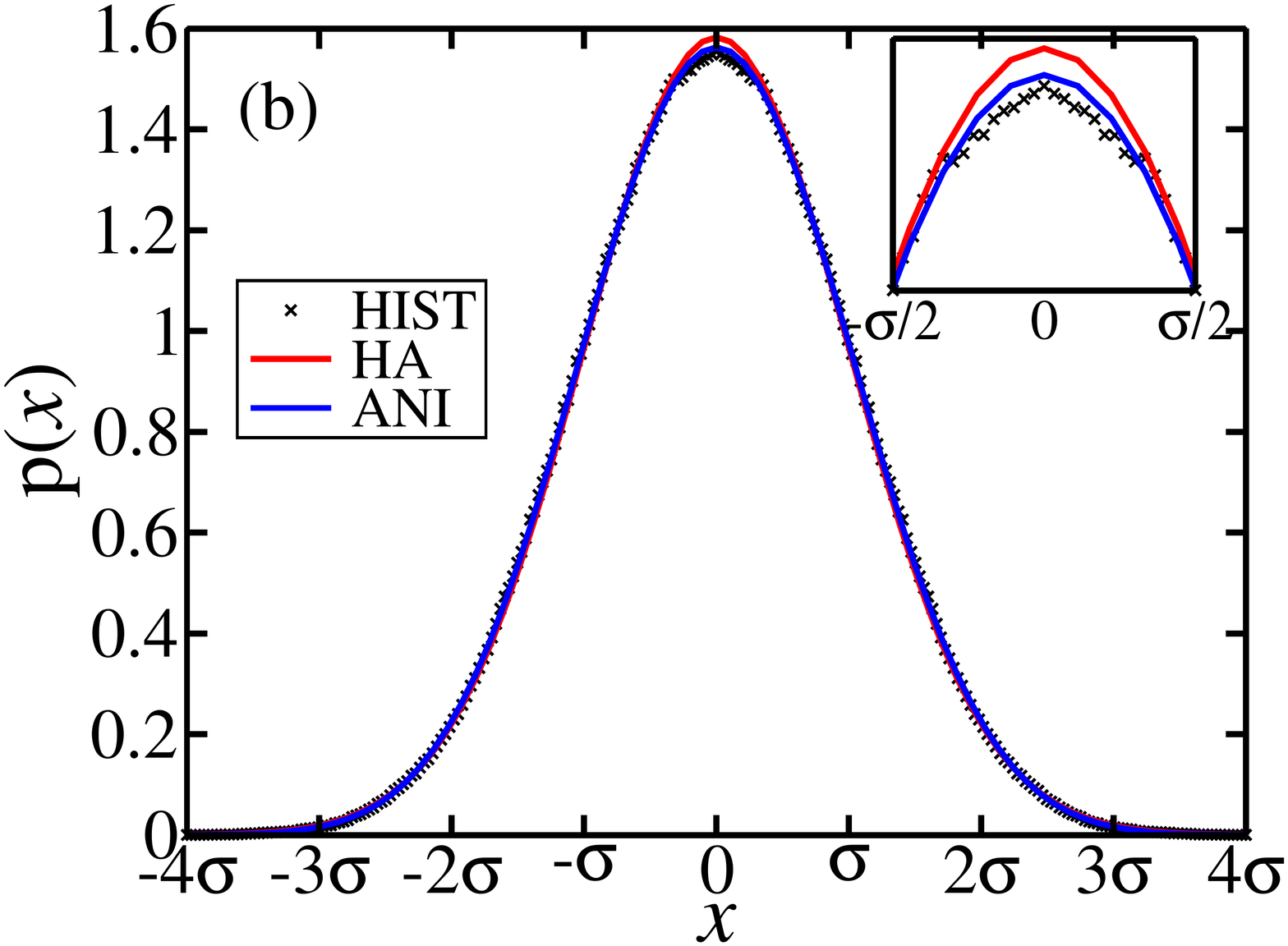}
  \caption{\label{fig:marginalx} Marginal probability distributions $p(x)$ of (a) FCC Al at T=500K and (b) BCC Ti at T=1200K. Crosses are histograms of the simulated data, red lines are fits to the harmonic model $p^h$ (Eq.~(\ref{eq:ph})), and blues lines are fits to the anharmonic model $p^a$ (Eq.~(\ref{eq:pa})).}
\end{figure*}

Table.~\ref{tab:solutions} compares the influence of anharmonicity in FCC Al with BCC Ti, and presents numerical values of the averages in Eq.~(\ref{eq:simultaneous}) and the solutions for $\sigma^2, a, b$ and entropy $S$. Anharmonicity tends to reduce the entropy for both FCC Al and BCC Ti, yet the reduction is insufficient to explain our entropy excess in BCC Ti.
Differences in the signs of the $a$ and $b$ parameters between FCC Al and BCC Ti imply opposite deviations of our harmonic model from the simulated distribution. In FCC Al, the simulated distribution is more narrow, with a higher probability at origin than in our harmonic model; in BCC Ti, the simulated distribution is broader and lower at the origin. Our anharmonic model captures these deviations, as shown in the marginal distributions $p(x)$ in Fig.~\ref{fig:marginalx}.

\begin{figure*}[htpb]
  \centering
  \includegraphics[width=.38\textwidth]{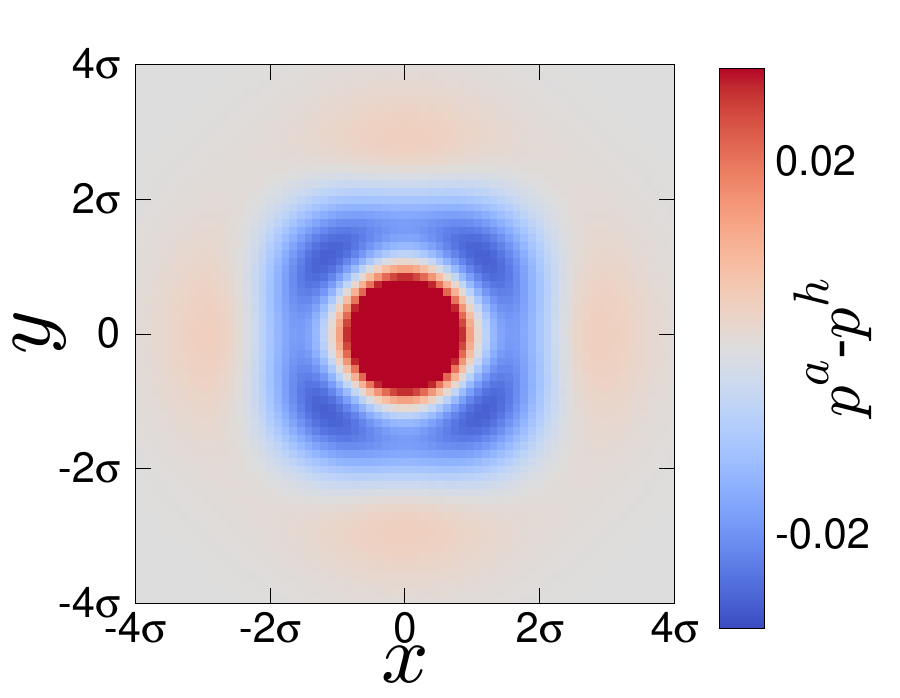}
  \includegraphics[width=.38\textwidth]{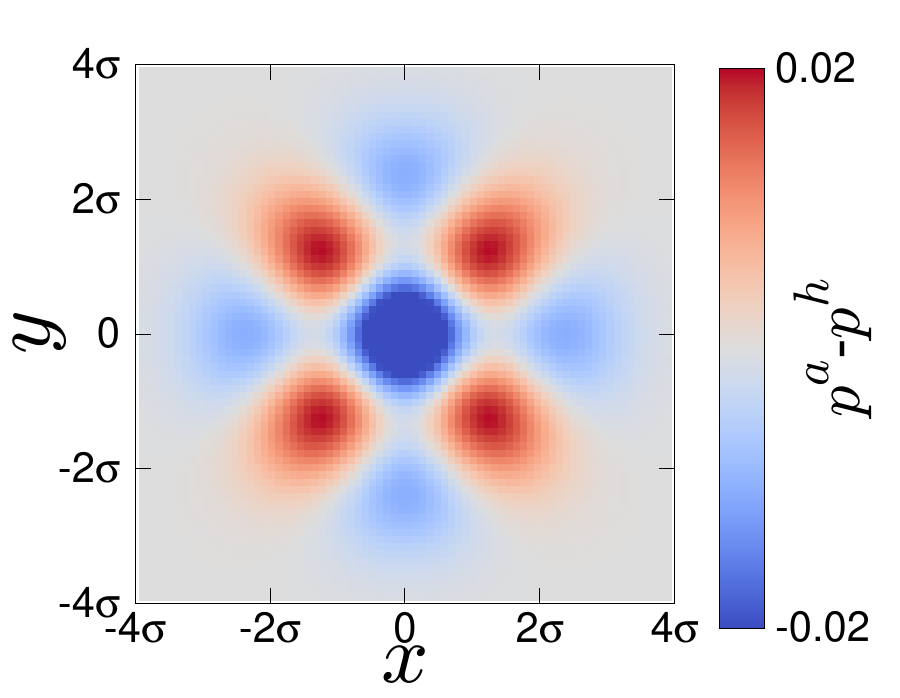}
  \caption{\label{fig:marginalxy}Difference of harmonic and anharmonic marginal probability distributions, $p^a(x,y)-p^h(x,y)$ for FCC Al (Left) and BCC Ti (Right).}
\end{figure*}

To examine the anisotropies, we plot the marginal distributions $p^a(x,y)$ in Fig.~\ref{fig:marginalxy}. In FCC Al, atomic displacements are reduced in the near-neighbor directions [110] and correspondingly enhanced in the [100] directions. In BCC the displacements are reduced in the nearest-neighbor directions [111] (not shown). Four such bonds project onto the [100] directions, while only two project onto [110], explaining the observed pattern. Overall, FCC Al is more isotropic than BCC Ti and hence has a smaller angular entropy correction.
 
\section{Conclusion}

We apply the information-theoretic entropy formula Eq.~(\ref{eq:S:info}) to evaluate the vibrational entropies of solids from the variance and covariance of atomic displacements. This approach generalizes prior work on the information-based entropy of liquids~\cite{GaoWidom2018,WidomGao2019,HuangGaoWidom2021}. In the case of liquids, the single atom entropy (ideal gas term) overestimates the entropy and must be corrected by removing the mutual information of the pair correlation functions. In the case of solids, the variance of individual atomic displacements can be measured through diffraction experiments that yield the Debye-Waller B-factor. Thus we find a crystallographic approach to estimate the thermodynamic entropy. However, as in the case of liquids, the one-body approximation overestimates the entropy by the information content of correlation functions, and we can improve the entropy estimate by including the covariance of atom pairs. This might be possible to achieve through diffraction experiments that measure the second-order thermal diffuse scattering~\cite{Warren1969}. It is easy to achieve through AIMD simulations of the atomic displacement covariance matrix, as we demonstrate in this paper for elemental Al and Na.

The method applies generally to solids, but the particular implementation given here relies on the accuracy of a Gaussian approximation to the distribution function. Hence it is most likely to work when the atomic displacements are small, and it is likely to fail in molecular solids where coherent bond rotations are present. Although we mainly demonstrated the method for elemental solids, it also holds in principle for complex crystalline and noncrystalline solids. We give an example of such an application for the MoNbTaW high entropy alloy.

The quasiharmonic method may be equally accurate and more efficient than our AIMD method when anharmonicity mainly enters through thermal expansion, but a simulation-based approach in principle includes additional anharmonic contributions. Doing so may require correlations beyond those captured by the Gaussian approximation (see Sec.~\ref{sec:anharmonic}). Our simulation-based approach seems most useful when the simulation has already been completed for other purposes. Then the entropy comes essentially for free on top of whatever other information was sought.

In some cases the quasiharmonic method cannot be applied due to the presence of imaginary frequency vibrational modes. The high temperature BCC phases of columns 3 and 4 of the periodic table exhibit such modes; they achieve mechanical stability only through their entropies. For elemental Ti, our AIMD method is capable of estimating the vibrational entropy, although the modes seem slightly softer than observed in experiment. We also point out the unexpected strong contribution to stability from the electronic entropy.

Our simulation approach based on the probability distribution is more flexible than the velocity-velocity correlation method because it does not rely on an underlying harmonic model, at least in the high temperature limit. Further, because it does not depend upon the dynamics, it can be used in conjunction with Monte Carlo simulation in addition to molecular dynamics. It requires only a single representative configuration provided the cell is sufficiently large, rather than demanding a long continuous trajectory.

\begin{acknowledgments}
  This work was supported by the Department of Energy under grant DE-SC0014506. We thank Don Nicholson for useful discussions and for alerting us to Reference~\cite{Nicholson2021}, and we thank Michael C. Gao for prior collaboration on the entropy of liquid metals. We benefited from computer time at the National Energy Research Scientific Computing Center (NERSC) using award ERCAP0015745.

\end{acknowledgments}

\bibliography{refs}
\end{document}